\documentclass{emulateapj}
\usepackage{graphicx}
\usepackage{epstopdf}
\usepackage{epsfig}

\newcommand{\Ang}{\mbox{ \AA}}     
\newcommand{\avg}[1]{\ensuremath{\langle #1 \rangle}}
\newcommand{\bma}{\begin{math}}
\newcommand{\ema}{\end{math}}
\newcommand{\beq}{\begin{equation}}
\newcommand{\eeq}{\end{equation}}
\newcommand{\beqa}{\begin{eqnarray}}
\newcommand{\eeqa}{\end{eqnarray}}
\newcommand{\bc}{\begin{center}}
\newcommand{\ec}{\end{center}} 
\newcommand{\bit}{\begin{itemize}}
\newcommand{\eit}{\end{itemize}}

\font\BFd=cmmib10
\font\BFt=cmmib10
\font\BFs=cmmib10 scaled 700
\font\BFss=cmmib10 scaled 500

\def\bbox#1{%
\relax\ifmmode
\mathchoice
{{\hbox{\BFd #1}}}
{{\hbox{\BFt #1}}}
{{\hbox{\BFs #1}}}
{{\hbox{\BFss #1}}}
\else \mbox{#1} \fi }

\begin{document}

%%%%%%%%%%%%%%%%%%%%%%%%%%%%%%%%%%%%%%%%%%%%%%%%%%%%%%%%%%%%%%%%%%%%%%%%%%%
% Front Material
%%%%%%%%%%%%%%%%%%%%%%%%%%%%%%%%%%%%%%%%%%%%%%%%%%%%%%%%%%%%%%%%%%%%%%%%%%%

%\twocolumn[%%% Begin front material

%\journalid{337}{15 January 1989}
%\articleid{11}{14}
 
\submitted{\today. To be submitted to \apj.} %@

\title{Quasar Proximity Zones and Patchy Reionization}
\author{Adam Lidz\altaffilmark{1}, Matthew McQuinn\altaffilmark{1}, Matias Zaldarriaga\altaffilmark{1,2}, Lars Hernquist\altaffilmark{1}, \& Suvendra Dutta\altaffilmark{1}}
\email{alidz@cfa.harvard.edu}

\altaffiltext{1}{Harvard-Smithsonian Center for Astrophysics, 60 Garden Street, Cambridge, MA 02138, USA}
\altaffiltext{2}{Jefferson Laboratory of Physics; Harvard University; Cambridge, MA 02138, USA}

\begin{abstract}
Lyman-alpha (Ly$\alpha$) forest absorption spectra towards quasars 
at $z \sim 6$ show regions of enhanced transmission close to their source. Several 
authors have argued that the apparently small sizes of these regions indicate that quasar 
ionization fronts at $z \gtrsim 6$ expand into a largely or partly neutral 
intergalactic medium (IGM). 
Assuming that the typical region in the IGM is reionized by $z \leq 6$, as is suggested by
Ly$\alpha$ forest observations, we argue that at {\em least} $50\%$ of the volume
of the IGM was reionized before the highest redshift quasars turned on. Further, even if the IGM
is as much as $\sim 50\%$ neutral at quasar turn-on, the quasars are likely born into large
galaxy-generated HII regions. The HII regions during reionization are themselves clustered, and
using radiative transfer simulations, we find that long skewers through the IGM towards quasar
progenitor halos pass entirely through ionized bubbles, even when the IGM is half neutral. These effects 
have been neglected in most
previous analyses of quasar proximity zones, which assumed a spatially {\em uniform} neutral
fraction. We model the subsequent ionization from a quasar, and construct mock Ly$\alpha$
forest spectra. Our mock absorption spectra are more sensitive
to the level of small-scale structure in the IGM than to the volume-averaged neutral 
fraction, and suggest
that existing proximity-zone size measurements are compatible with a fully 
ionized IGM. However, we mention
several improvements in our modeling that are necessary to make more definitive conclusions.
\end{abstract}

\keywords{cosmology: theory -- reionization -- intergalactic medium -- large scale
structure of universe}
%]%%% End front material

%%%%%%%%%%%%%%%%%%%%%%%%%%%%%%%%%%%%%%%%%%%%%%%%%%%%%%%%%%%%%%%%%%%%%%%%%%%
\section{Introduction} \label{sec:intro}
%%%%%%%%%%%%%%%%%%%%%%%%%%%%%%%%%%%%%%%%%%%%%%%%%%%%%%%%%%%%%%%%%%%%%%%%%%%

The Epoch of Reionization (EoR), when HII regions grow around galaxies
and/or quasars, eventually overlap and fill the entire IGM, is 
fundamental to our understanding of cosmological structure
formation. Detailed observations of the EoR will characterize the
nature of the first luminous sources in the Universe, describe their
impact on the surrounding IGM, and fill in a significant gap in our
knowledge of the history of the Universe.  Prospects for observational
advances are bright: 21cm observations (e.g. Madau et al. 1997,
Zaldarriaga et al. 2004, for a review see Furlanetto et al. 2006a),
improved measurements of polarization of the cosmic microwave
background (CMB) (Zaldarriaga 1997, Kaplinghat et al. 2003),
small-scale CMB fluctuations (e.g. Zahn et al. 2005, McQuinn et
al. 2005), increasingly deep narrow-band Ly-$\alpha$ surveys
(e.g. Haiman \& Spaans 1999, Barton et al. 2004, Furlanetto 2006b, 
McQuinn et al. in prep.), optical 
afterglow spectra of gamma ray
bursts (GRBs) (Barkana \& Loeb 2004), quasar absorption spectra (Fan
et al. 2006), and other probes, promise a wealth of new data in the
near future.

What have we learned from existing observations?  This is, of course,
an intrinsically interesting question, but it is also an important one
for directing the design of future surveys and experiments. For
example, current constraints can provide important guidance regarding
the optimal target redshift range for future reionization surveys.

Inferences about the duration of the EoR come from measurements of
anisotropies in the polarization of the CMB (Page et al. 2006),
narrow-band surveys for Ly-$\alpha$ emitters (e.g. Malhotra \& Rhoads
2005), the optical afterglow of a $z = 6.3$ GRB (Totani et al. 2006),
and particularly the absorption spectra of high redshift quasars
(e.g. Fan et al. 2006).  While valuable and exciting, these
observations have yielded constraints on the EoR that are generally
weak and subtle to interpret.  The central value for the electron
scattering optical depth from WMAP favors reionization activity at $z
\gtrsim 11$, but, at the $1-\sigma$ level, current limits are
consistent with rapid reionization ending near $z \sim 6$ (Page et
al. 2006).  Indeed, at $\sim 3-\sigma$ the data are consistent with no
reionization whatsoever (Page et al. 2006).

Malhotra \& Rhoads (2005) have constrained the ionized volume fraction
in the IGM by counting the abundance of Ly-$\alpha$ emitters at $z =
6.5$, and requiring a minimum ionized volume around the sources to
avoid attenuating their Ly-$\alpha$ photons. In this manner, Malhotra
\& Rhoads quote an upper limit on the neutral volume fraction of the
IGM, $X_{\rm HI} \lesssim 0.5$.  In reality, the sources likely reside
in much larger ionized regions (e.g. Furlanetto et al. 2006b), and
this constraint should be tightened with more detailed modeling.
For example, Dijkstra at al. (2006) find that these observations imply
$X_{\rm HI} \lesssim 0.2$.

The optical afterglow spectra of $z \gtrsim 6$ GRBs can, in principle,
be used to search for damping-wing absorption redward of Ly-$\alpha$,
a signature of a largely neutral IGM (Miralda-Escud\'e 1998, Barkana
\& Loeb 2004). However, most GRB optical afterglows show evidence for
damped Lyman-alpha absorbers associated with their host galaxies,
which are problematic to distinguish from a largely neutral IGM
(Totani et al. 2006). Even so, Totani et al. (2006) use the Ly-$\beta$
region of a $z=6.3$ afterglow to argue against a largely neutral IGM,
giving a constraint of $X_{\rm HI} \lesssim 0.6$.

Presently, the most detailed information on the state of the IGM at $z
\sim 6$ comes from Ly-$\alpha$ forest absorption towards high
redshift quasars (e.g. Fan et al. 2006).  It is difficult to derive
constraints on the ionization state of the IGM from the spectra of
these quasars, owing to the large absorption cross section for
Ly-$\alpha$ photons. Indeed, a highly ionized IGM ($X_{\rm HI} \gtrsim
10^{-4}$) results in complete absorption in a $z \sim 6$ Ly-$\alpha$
forest spectrum (Gunn \& Peterson 1965). In spite of this intrinsic
complication, a little ingenuity has led to substantial progress.

For example, one can measure absorption in the Ly-$\beta$ and
Ly-$\gamma$ troughs of a quasar spectrum. Owing to their weaker
absorption cross sections, when these transitions are saturated they
imply tighter constraints on the ionization state of the IGM than the
optical depth in Ly-$\alpha$.  This approach has been used to suggest
that the IGM is evolving rapidly near $z \sim 6$ (e.g. Fan et
al. 2002, Cen \& McDonald 2002, Lidz et al. 2002, Fan et
al. 2006). The constraints are, however, consistent with a mostly
ionized IGM.  Furthermore, the transmission is influenced strongly by
rare underdense regions at high redshift, and so the conclusions
depend sensitively on the probability distribution of gas in the IGM
(Oh \& Furlanetto 2005, Becker et al. 2006a).  Another interesting
statistic is to consider how much the absorption, averaged over large
stretches of spectra, varies from sightline-to-sightline.  Close to
and during reionization, the ultraviolet radiation background should
fluctuate strongly (e.g. Zuo 1992, Wyithe \& Loeb 2005a) and
potentially increase the sightline-to-sightline scatter in the mean
absorption.  However, current measurements are broadly consistent with
density fluctuations alone (Lidz et al. 2006a, Liu et al. 2006).

It is also possible to use a metal line tracer of the ionization state
of the IGM, such as OI, which conveniently lies redward of Ly-$\alpha$
and has an ionization potential similar to that of hydrogen (Oh
2002). In fact, high-resolution Keck spectra of the $z \sim 6$ SDSS
quasars do reveal some OI lines, with $4$ out of $6$ detected systems
lying towards the highest redshift quasar known (Becker et al. 2006b).
Interestingly, some of the OI systems are nearby regions that show
transmission in the Ly$\alpha$ and Ly$\beta$ forests of this quasar
(Becker et al. 2006b).  The interpretation of these observations is
unclear: the OI systems might reflect dense clumps of neutral gas in a
highly ionized IGM, or instead could indicate inhomogeneous metal
pollution in a more neutral IGM.

Finally, the tightest constraints claimed on the ionization state of
the $z \sim 6$ IGM come from measurements of the proximity regions
around $z \sim 6$ quasars.  Several authors, starting with Wyithe
\& Loeb (2004), have argued that these regions are small,
indicating that quasar ionization fronts are expanding into a largely
neutral IGM.  Mesinger \& Haiman (2004) claim to detect the edge of a
quasar ionization front around one of the SDSS $z \sim 6$ quasars,
using additional leverage from the Ly$\beta$ region, and suggest that
the neutral fraction is $X_{\rm HI} \gtrsim 0.25$ at $z \sim 6$. These authors'
constraint comes not so much from the apparent size of the proximity zone, but from the
detailed radial dependence of the transmission and from a stretch
of spectrum with Ly$\beta$ transmission and no corresponding Ly$\alpha$ transmission.
They attribute this to damping wing absorption, a signature of a partly neutral IGM.
Oh \& Furlanetto (2005), however, argue that such stretches occur at high redshift even
when the IGM is highly ionized and may not indicate damping wing absorption.
Fan
et al. (2006) emphasize caution in interpreting proximity region
measurements, but observe rapid evolution in the sizes of quasar
proximity regions from $z = 5.7-6.3$, arguing for a correspondingly
rapid evolution in the neutral fraction. Their final constraint is
based on only the evolution of the proximity region size, which they
argue reflects the change in the neutral fraction.  Consequently, they
quote a significantly more conservative limit than previous authors,
requiring a volume-weighted neutral fraction of only $X_{\rm HI}
\gtrsim 10^{-3}$.

Recently, more detailed proximity zone calculations have been
performed.  Bolton \& Haehnelt (2006) studied quasar transmission
carefully using 1D radiative transfer calculations, and determined
that it is difficult in general to distinguish highly ionized and
mostly neutral models with proximity zone measurements. Maselli et
al. (2006) came to a similar conclusion.  Mesinger \& Haiman (2006),
however, argue that detailed fitting of the transmission pdf in the
quasar proximity zones favors a partly neutral IGM, with a lower limit
of $x_{\rm HI} \gtrsim 0.033$, and a considerably larger preferred value.

In each of these studies, the authors have assumed that quasar
ionization fronts expand into a uniformly ionized surrounding IGM,
with some low level, yet {\it homogeneous} background ionization. If
the $z \gtrsim 6$ quasar spectra truly probe the pre-reionization
epoch, then this is likely a very poor approximation.  The
pre-reionization IGM should resemble swiss cheese (Loeb 2006), with
large HII regions forming around clustered galaxies embedded in a
surrounding neutral IGM (e.g. Sokasian et al. 2003, Ciardi et
al. 2003, Furlanetto et al. 2004a,c, Iliev et al. 2006, Zahn et
al. 2006, McQuinn et al. 2006a, Trac \& Cen 2006).  Naively, this
complicates distinguishing partly neutral and highly ionized models on
the basis of quasar proximity zones.  Quasar ionization fronts may
extend further along sightlines that traverse several ionized HII
regions, and be more limited along other directions that traverse
several neutral patches.  Moreover, transmission at the edge of the
proximity zone may be related to background galaxies rather than the
quasar itself, making it still harder to locate the `edge' of the
proximity zone using absorption spectra (see also Wyithe
\& Loeb 2006a).  Our present paper extends these earlier works, and
focuses on how `patchy reionization' impacts quasar proximity zones.

The outline of our paper and our basic line of argument is as
follows. Given that most of the volume of the IGM appears to be highly
ionized by $z \lesssim 6$, we argue that reionization is unlikely
rapid enough for the IGM to be mostly neutral when the highest
redshift quasars observed turned on (\S \ref{sec:ic}). From these
considerations, we suggest that the IGM is {\em at least} $50\%$
ionized at quasar turn-on.  In \S \ref{sec:3d_rt}, we describe 3D
radiative transfer calculations for plausible partly neutral models,
detailing the initial ionization state of the IGM.  Here we argue that
the highest redshift quasars are born into large HII regions, even if
as much as $50\%$ of the volume of the IGM is neutral. Yu \& Lu (2005)
previously argued that overdense $z \sim 6$ quasar environments should
reionize before typical regions, but here we examine the consequences
of this in more detail, and arrive ultimately at somewhat different
conclusions.  Even in this maximally ($50\%$) neutral scenario, we
find long skewers towards quasar progenitor halos which pass entirely,
or predominantly, through ionized bubbles.

Using the initial ionization field from our 3D calculations as input,
we perform (more detailed) 1D radiative transfer calculations,
describing the subsequent propagation of quasar ionization fronts (\S
\ref{sec:1d_rt}).  Here we find that quasar front extents, in our
patchy reionization models, depend sensitively on the long ionized
pathways created by surrounding galaxies {\it before} the quasar is
born. Since quasars are typically born into large HII regions, the
fronts tend to extend {\em further} in patchy reionization models
than in models with a uniform IGM of the same neutral fraction, although with
significant sightline-to-sightline variation.  In \S
\ref{sec:mock_spectra} we construct mock absorption spectra and show,
in agreement with previous authors (Bolton \& Haehnelt 2006, Maselli
et al. 2006), that it is difficult to accurately recover the position
of a quasar front, and that estimates of the front position from
absorption spectra are generally {\em underestimates}.  This is a
consequence of the typically high Ly$\alpha$ opacity at the edge of a
$z \sim 6$ quasar front. 
We suggest, however, an alternative algorithm
for finding front positions from absorption spectra. We then estimate
the importance of unresolved (in our numerical simulations) small
scale structure on our mock absorption spectra.  Our results are more
sensitive to the highly uncertain level of small scale structure in
the IGM than to the volume-weighted ionization fraction.  Finally, in
\S \ref{sec:conclusions} we conclude and discuss possible future research
directions.

Throughout, we assume a flat, $\Lambda$CDM cosmology parameterized by:
$\Omega_m=0.3$, $\Omega_\Lambda=0.7$, $\Omega_b=0.04$, $H_0 = 100 h$
km/s/Mpc with $h=0.7$, and a scale-invariant primordial power spectrum
with $n=1$, normalized to $\sigma_8(z=0)=0.9$. Our adopted value for
$\sigma_8$ is a little larger than the central value favored by
$3$-year constraints from WMAP (Spergel et al. 2006), but
uncertainties in the cosmological model are sub-dominant to other
aspects of our theoretical modeling, which we detail subsequently.

%%%%%%%%%%%%%%%%%%%%%%%%%%%%%%%%%%%%%%%%%%%%%%%%%%%%%%%%%%%%%%%%%%%%%%%%%%%%
\section{Initial Conditions} \label{sec:ic}
%%%%%%%%%%%%%%%%%%%%%%%%%%%%%%%%%%%%%%%%%%%%%%%%%%%%%%%%%%%%%%%%%%%%%%%%%%%%

Now, we consider the question: what should we expect for the
ionization state of the gas around $z \gtrsim 6$ quasars when
they turn on?  In this section, we address this issue using a rough
analytic model; in the subsequent section we describe 3D radiative
transfer calculations which examine this in more detail.

First, recall that constraints from $z \lesssim 6$ quasar spectra
suggest that the IGM is highly ionized by $z \lesssim 6$ (e.g. Fan et
al. 2002), although see \S \ref{sec:conclusions} for a critical
discussion.  To date, the highest redshift quasar observed is at $z_q =
6.42$. If this source turns on over roughly a Salpeter time ($t_s \sim 4
\times 10^7$ yrs) before being observed, then its turn-on
redshift is $z_{\rm on} \sim 6.66$.  For the surrounding IGM to be
mostly neutral at quasar birth, yet highly ionized by $z \lesssim 6$,
reionization must proceed extremely rapidly, occurring over a short
redshift span of $\Delta z \lesssim 0.7$ or $\sim 10^8$ yrs.
Obviously, reionization must occur even more rapidly if measurements
truly indicate that other quasars at slightly lower redshifts with
$z_q \gtrsim 6$, (e.g. the $z_q = 6.28$ quasar SDSS J1030+0524), are
also born when the surrounding IGM is highly neutral.

Second, the $z \gtrsim 6$ quasars are thought to reside in very rare
and massive host halos (Fan et al. 2001, Li et al. 2006), and hence
the surrounding IGM is likely overdense out to large scales around the
quasars (Loeb \& Eisenstein 1995, Barkana 2004, Faucher-Gigu\`ere et
al. 2007). These regions should reionize before typical ones since
halo collapse and galaxy formation are expected to occur earlier
in large-scale overdensities (e.g. Barkana \& Loeb 2004).  The
abundance of ionizing sources in an overdense region at a given time
should resemble the abundance of sources in a typical region at {\it a
later time}.  Even if typical regions in the IGM are neutral when
quasars turn on, the same may not be true of the overdense
environments where quasars reside (see also Yu \& Lu 2005).
Recombinations are also more efficient in overdense regions -- this
could offset the tendency for overdensities to reionize first, but
this is unlikely to remove the trend.  This is because galaxy
formation is more sensitive to large scale overdensity than the
recombination rate -- indeed, the abundance of high mass halos is
exponentially sensitive to the large scale overdensity (Barkana \&
Loeb 2004, Furlanetto \& Oh 2005, Wyithe \& Loeb 2006b).

\subsection{Globally-Averaged Ionization Fractions}
\label{sec:xglobal}

In this section we illustrate these effects using simple analytic
estimates.  We perform these calculations with the analytic model of
Furlanetto et al. (2004).  In its simplest incarnation, this method
assumes that a galaxy of mass $M_{\rm gal}$ can ionize a surrounding
mass in the IGM of $M_{\rm ion} = \zeta M_{\rm gal}$.  With this
assumption, one can show that the average ionization fraction of gas
in a region of radius $R$ and linear overdensity $\delta_R$ is given
by:
\beqa
\avg{x_i|\delta_R, R} = \zeta (f_{\rm coll}|M_{\rm min}, R, \delta_R).
\label{eq:xbar_delta}
\eeqa
Here, $(f_{\rm coll}|M_{\rm min}, R, \delta_R)$ is the fraction of
mass in halos with mass larger than $M_{\rm min}$ in a region of
over-density $\delta_R$ and radius $R$.  In this equation, the impact
of recombinations is absorbed into the parameter $\zeta$.  For our
simple estimates below we hence ignore the recombination-rate 
enhancement in over-dense regions.

The collapse fraction in an overdense region is given by:
\beqa
(f_{\rm coll}|M_{\rm min}, \delta_c, M, \delta_M) = \rm{erfc}\left[\frac{\delta_c - \delta_M}
{\sqrt{2 \sigma^2_{\rm min} - 2 \sigma^2_M}}\right] \, ,
\label{eq:fcoll_delta}
\eeqa
where $\delta_c$ is the critical overdensity for collapse, $M$
indicates the mass scale corresponding to the spatial scale $R$, and
$\sigma^2_{\rm min}$ and $\sigma^2_M$ are the variance of the
linear density field smoothed on mass scales $M_{\rm min}$ and $M$,
respectively.  The {\it global} collapse fraction follows from this
expression in the limit $\delta_M \rightarrow 0$, $\sigma^2_M
\rightarrow 0$. This equation implies that overdense regions will have
larger collapse fractions than regions at the mean density and, under
the assumptions of Equation (\ref{eq:xbar_delta}), will be ionized
earlier.

Consder, first, the redshift evolution of the globally-averaged
ionization fraction, by taking $\delta_R \rightarrow 0$ and $R
\rightarrow \infty$ in Equation (\ref{eq:xbar_delta}).  The precise
duration of reionization depends sensitively on the nature of the
sources producing ionizing photons at early times, which is highly
uncertain. If rare, yet efficient sources reionize the IGM then the
collapse and ionization fractions should grow rapidly with redshift
(Equations \ref{eq:xbar_delta} and \ref{eq:fcoll_delta}). Furthermore,
the duration of the reionization epoch depends on the efficiency of
thermal feedback, the time taken to photo-evaporate mini-halos in the
IGM, as well as the detailed properties and evolution of the cosmic
star-formation rate and the fraction of ionizing photons escaping into
the IGM.  In our model, all of these details are subsumed into a
single, redshift-independent parameter, $\zeta$ (Equation
\ref{eq:xbar_delta}).

Nevertheless, we can roughly gauge the range of possibilities by
varying the parameter $M_{\rm min}$ in Equation (\ref{eq:fcoll_delta}),
in each case normalizing $\zeta$ to match $\avg{x_i}=1$ at $z=6$. Of
course, reionization may finish at significantly higher redshift
(i.e., $\avg{x_i}=1$ is reached at $z > 6$); our goal here is to find
the maximally neutral case at quasar turn-on provided that the entire
volume is reionized by $z \lesssim 6$. We consider a wide range of
values for $M_{\rm min}$.  On the low end, we adopt $M_{\rm min} =
M_{\rm cool}$, and on the high end we take $M_{\rm min} = 10^3 M_{\rm
cool}$. Here, $M_{\rm cool}$ denotes the dark matter halo mass
corresponding to a virial temperature of $10^4$ K ($M_{\rm cool} = 1.3
\times 10^8 M_\odot$ at $z \sim 7$, e.g. Barkana \& Loeb 2001), above
which atomic line cooling is efficient, allowing gas to cool, condense
to form stars, and produce ionizing photons. The higher minimum source
mass scenarios approximate models where photo-heating has limited the
efficiency of star-formation in small mass halos (Thoul \& Weinberg
1996, Navarro \& Steinmetz 1997, Dijkstra et al. 2004), and models in
which supernova winds suppress star-formation in low mass halos
(e.g. Springel \& Hernquist 2003). We note that our highest minimum
source mass model ($M_{\rm min} = 10^3 M_{\rm cool}$) is rather
extreme, considerably larger than the suppression masses suggested by
the above studies. We consider this case anyway to illustrate a
plausible upper limit.

\begin{figure}
\bc
\includegraphics[width=9.2cm]{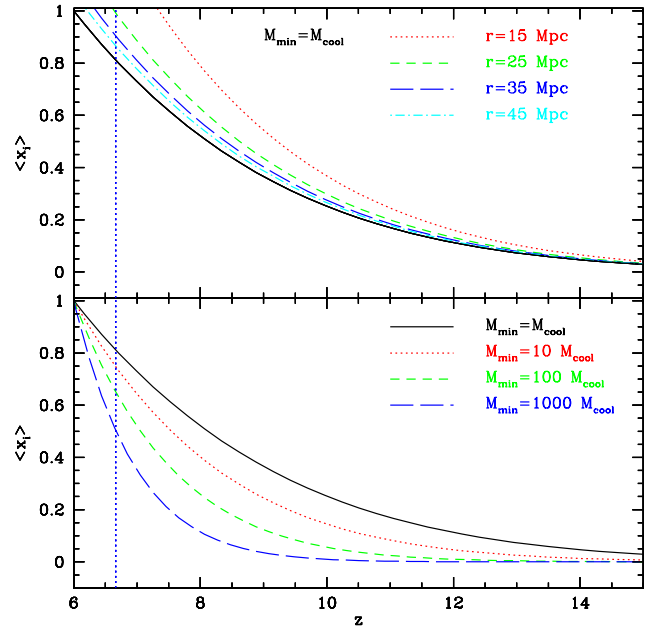}
\caption{Redshift evolution of the volume-averaged ionization fraction. {\em Top panel:} The black
solid line shows the redshift evolution of the ionization fraction in a typical
region of the IGM, with the ionizing source efficiency calibrated so that $\avg{x_i} = 1$
at $z=6$, assuming all halos down to $M_{\rm cool}$ contribute ionizing photons with equal efficiency.
The other curves indicate the average ionization fraction in spherical regions of different size, 
each with a massive quasar host (here $M_{\rm dm, host} = 10^{13} M_\odot$) at the center. 
The vertical blue dotted line indicates a plausible turn-on time for a $z=6.42$ quasar.
The curves illustrate that: i) the process of reionization takes a sufficiently long time that, even if
it completes at $z=6$, the typical location in the IGM is already $\sim 80\%$
ionized (in this model) when the $z=6.42$ quasar turned on; and ii) provided that quasar host halos
reside in highly overdense regions, their surroundings will reionize {\em earlier than the 
typical region in the IGM.} 
{\em Bottom panel:} The redshift evolution of the volume-averaged ionization fraction for a typical
region as a function of the minimum host halo mass.
If rare, very efficient, sources produce most of the 
ionizing photons, reionization occurs more rapidly than if highly abundant yet less efficient sources 
produce most of the ionizing photons. In each model, more than $50\%$ of the IGM volume is ionized
at quasar turn-on.
}
\label{fig:xbar_enhanced}
\ec
\end{figure}

We show the redshift evolution of the globally-averaged ionization
fraction in the bottom panel of Figure \ref{fig:xbar_enhanced}. The
first qualitative feature apparent from the Figure is that
reionization is generally quite extended: for example, in the cooling
mass model the average ionization fraction is $\avg{x_i}=0.1, 0.5,
0.7, 1.$ at $z \sim 12, 8, 7$, and $z \sim 6$ respectively.  If the
highest redshift quasar observed thus far, at $z_q=6.42$, turns on at
$z_{\rm on} \sim 6.66$ as motivated above, the average ionization
fraction in this model is $\sim 80\%$ at turn-on.  The process is less
extended if the sources are rare, yet very efficient.  For example, in
the $M_{\rm min} = 10^2 M_{\rm cool}$ case, the IGM is only $\sim
60\%$ ionized at quasar turn on. However, even in the very extreme
$M_{\rm min} = 10^3 M_{\rm cool}$ scenario, $50 \%$ of the IGM is
ionized by quasar turn on at $z \sim 6.66$. The ionization fraction
evolves more rapidly in the high minimum mass models, since here the
host halos are still on the exponential tail of the mass-function near
$z \sim 6$, and hence their abundance evolves quickly with
redshift. Note that we likely over-estimate this effect, since
we use Press-Schechter (1974) theory for simplicity: the halo
abundance from recent high redshift numerical simulations more closely
matches the Sheth-Tormen (1999) mass function, indicating that
Press-Schechter theory underestimates the abundance of rare halos
(Reed et al. 2003, Heitmann et al. 2006, Zahn et al. 2006, Lukic et
al. 2007, although see Trac \& Cen 2006). Note that although our constraint argues
that the IGM is at least somewhat ionized when the highest redshift quasar observed
turns on, it is not in conflict with the formal limits from
Wyithe \& Loeb (2004),
Wyithe et al. (2005a), and Mesinger \& Haiman (2004, 2006).

Furthermore, we have taken $\zeta$ as a free parameter, simply fixing
it to match $\avg{x_i}=1$ at $z \sim 6$. In fact, if all of the
ionizing photons are indeed produced by the very rare halos with $M
\gtrsim M_{\rm min} = 10^3 M_{\rm cool}$, the sources need to be
exceedingly efficient to reionize the IGM by $z \lesssim 6$.
Specifically, we estimate the number of ionizing photons per stellar
baryon required in this model to achieve $\avg{x_i}=1$ by $z=6$, using
Equations (82) and (83) of Furlanetto et al.  (2006a). We adopt
typical values of three recombinations per ionized hydrogen atom, an
escape fraction of $f_{\rm esc} = 0.1$, and a star-formation
efficiency of $f_\star = 0.1$, and find that $N_\gamma \sim 70,000$
ionizing photons per baryon are required in this scenario. This
ionizing efficiency is substantially larger than the value expected
for a Salpeter IMF, $N_\gamma \sim 4,000$ (e.g. Cohn \& Chang 2006),
although it is achievable if the IMF is very top-heavy (e.g. Bromm et
al. 2001) even at $z \sim 7$ in spite of apparently wide-spread metal
enrichment by $z \sim 6$ (Ryan-Weber et al. 2006).  

Finally, in \S \ref{sec:conclusions} we argue that it is conceivable
that $\avg{x_i}=1$ is achieved later than $z=6$ -- reionization might
end at as low a redshift as $z \sim 5.5$. Even if reionization completes
only by $z=5.5$ we find that the neutral
fraction at quasar turn-on is less than $50\%$ in all of our models except our extreme
$M_{\rm min} = 10^3 M_{\rm cool}$ model. Moreover, reionization is likely
more extended than in our model -- for example, thermal feedback and mini-halos
can extend the duration of reionization compared to our simple
predictions (e.g. Haiman \& Holder 2003, McQuinn et al. 2006a).
Therefore, our
simple estimates indicate that the IGM is unlikely as much as $50\%$
neutral at quasar turn-on and is more likely at least $70\%$ ionized
by this time.
These rough estimates are inevitably model dependent, but are in
accord with observational constraints from Ly$\alpha$ emitters
(Malhotra \& Rhoads 2005, Haiman \& Cen 2005, Dijkstra et al. 2006), and GRB optical
afterglow spectra (Totani et al. 2006) which coincidentally give
similar upper limits for the neutral fraction near the plausible
turn-on redshifts of the $z \gtrsim 6$ quasars.

\subsection{The Ionization Field Around Quasar Progenitor Halos}
\label{sec:xqso}

The above calculations apply to {\em typical} regions in the IGM. We
now extend our analysis to consider the ionization of the overdense
environments expected around high redshift quasars. We will compute
the probability distribution of the ionized fraction
spherically-averaged around plausible quasar host halos, prior to
quasar turn-on. Here, our calculation is similar to previous work by Yu
\& Lu (2005), but it differs in detail.  We first consider the
linear overdensity profiles around the $z \sim 6$ quasars following
Loeb \& Eisenstein (1995) and Barkana (2004). We will subsequently
insert the linear density profile into Equation (\ref{eq:xbar_delta}).
The first quantity of interest is the cross-correlation coefficient,
$\rho_{\rm r, R}$, between density fluctuations at a single point when
smoothed on two different scales, $r$ and $R$. The correlation
coefficient is related to the co-variance, $\sigma^2_{\rm r, R}$, and
individual variances, $\sigma^2_r$, $\sigma^2_R$, by the relation
$\rho_{\rm r, R} = \sigma^2_{\rm r, R}/\sqrt{\sigma^2_r
\sigma^2_R}$. The covariance can be expressed as an integral over the
linear density power spectrum, $P(k)$, through the relation
\beqa
\sigma^2_{\rm r, R} = \int \frac{d^3k}{(2\pi)^3} W(kr) W(kR) P(k).
\label{eq:covar}
\eeqa

If the window functions in this equation are (spherically symmetric) 
top-hat filters in $k$-space then the co-variance
is precisely equivalent to the variance on scale $R$ -- i.e., $\sigma^2_{\rm r, R} = \sigma^2_R$,
and the correlation coefficient is given by $\rho_{\rm r, R} = \sigma_R/\sigma_r$.
The co-variance will be a little different if the window functions are top-hats in real space. In spite of
this, for simplicity we adopt the sharp $k$-space correlation coefficient even though we calculate
$\sigma_R$ and $\sigma_r$ using a real-space top hat, as is frequently done in Press-Schechter type calculations.

In what follows, we ignore the `cloud-in-cloud' problem of Press-Schechter
theory, and do not include an `absorbing barrier' in our calculation (Bond et
al. 1991). Barkana (2004) constructs a more elaborate model for the
initial overdensity profile around massive halos that is consistent
with extended Press-Schechter theory. However, in the limit of the very rare
halos we consider here, Barkana (2004) shows that this more elaborate model
reduces to our present one, which we hence adopt for simplicity.
Assuming the density field is a bi-variate Gaussian with the correlation coefficient given 
above, we can immediately
write down an expression for the desired conditional probability distribution: we would
like to know the differential probability that a point is at a linear overdensity $\tilde{\delta}_R$
when smoothed on a scale $R$ at some redshift $z_1$, given that it is at linear overdensity 
$\tilde{\delta}_c$ when
smoothed on a smaller scale, $r$ at redshift $z_2$. This expression is:
\beqa
P(\delta_R, \sigma_R, z_1 | \delta_c, \sigma_r, z_2)  = \frac{1}{\sqrt{2 \pi}} \frac{\sigma_r}
{\sigma_R \left[\sigma^2_r - \sigma^2_R \right]^{1/2}} \nonumber \\ 
 \times \rm{exp}\left[-\frac{(\delta_c - \delta_R)^2}{2 \sigma^2_r - 2 \sigma^2_R}\right] 
\rm{exp}\left[\frac{\delta^2_c}{2 \sigma^2_r} - \frac{\delta^2_R}{2 \sigma^2_R}\right].
\label{eq:cond_prob}
\eeqa

In this equation, all quantities are linearly-extrapolated to the
present day. Hence, if $\tilde{\delta}_R$ is the linear overdensity at
smoothing scale $R$ and redshift $z_1$, and $D(z_1)$ is the linear
growth factor normalized to unity today, then $\delta_R =
\tilde{\delta}_R/D(z_1)$ is the linear overdensity on scale $R$ today.
If we take $\delta_c$ to be the critical overdensity for collapse
scaled to the present day, then this equation tells us the conditional
probability that a region will have a large scale overdensity,
$\delta_R$ on scale $R$, given that the region contains a massive halo
-- i.e., the region reaches the collapse threshold, $\delta_c$, at a
smaller smoothing scale, $r$. Combining Equations (\ref{eq:xbar_delta}),
(\ref{eq:fcoll_delta}), and (\ref{eq:cond_prob}), and computing the
Jacobian $|dx_i/d\delta_R|^{-1}$, we can determine the desired
ionization probability distribution.  The ionization fraction,
averaged over an ensemble of quasar host halos, just follows from
determining the mean of the resulting ionization probability
distribution. Note that in this calculation we neglect sources {\em
outside} the region of interest, but since we are interested in rather
large smoothing scales, this is probably not too poor an
approximation.

Our results for the {\em average} ionization in regions that will host
a massive quasar host halo at $z_2=z_q=6.42$ are shown in the top
panel of Figure \ref{fig:xbar_enhanced}. Here we assume that $z
\gtrsim 6$ quasars reside in $10^{13} M_\odot$ halos as in Haiman \& Loeb (2001) and 
Li et al. (2006), although this choice is uncertain.  Our simulation results
in subsequent sections assume quasars reside in slightly less massive
and more common halos.  For the present calculation we adopt our model
in which all host halos down to $M_{\rm cool}$ contain ionizing
sources and contribute to reionization.  Figure
\ref{fig:xbar_enhanced} demonstrates that a significant volume of gas
around quasar progenitors is generally reionized well before
reionization completes in a typical region of the IGM.  Indeed, $\sim
100\%$ of the volume in a sphere of radius $15$ co-moving Mpc centered
on the $z=6.42$ quasar progenitor is ionized in our model, on average,
by $z \gtrsim 7$, $\sim 50$ Myr before the quasar turns on.  As one
averages over progressively larger volumes, the mean interior
overdensity approaches the cosmic average, and the ionization
approaches its cosmic mean value. Indeed, the figure indicates that
spheres of radius $35-45$ Mpc/$h$ appear to be essentially
representative samples. Given that the purported proximity zone size
is $\sim 40$ co-moving Mpc/$h$ (e.g. Wyithe et al. 2005a), one might
conclude that this bias is hence unimportant for interpreting quasar
proximity zone measurements. In fact, we will show in the next section
that 1D skewers through the IGM towards quasar host halos can pass
entirely through HII bubbles out to much larger distances. Therefore
the tendency for quasars to be born into large HII bubbles is in fact
important for interpreting the proximity zones in $z \sim 6$ quasar
absorption spectra.

\begin{figure}
\bc
\includegraphics[width=9.2cm]{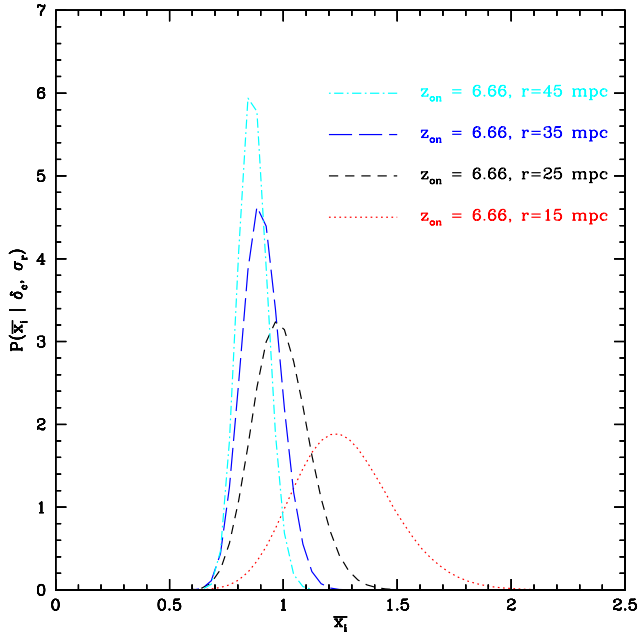}
\caption{Probability distribution for the spherically-averaged ionization fraction when the $z=6.42$ quasar 
turns on in our model for several different smoothing scales. The curves show pdfs for the fraction of the 
volume that is ionized in regions of different size
centered on the quasar host halo. A fraction of $x_i \geq 1$ simply indicates that there are enough
photons to ionize the entire volume of the region in question.
}
\label{fig:pdf_xbar}
\ec
\end{figure} 

We can also examine the full probability distribution of the
spherically-averaged ionization fraction around quasar host halos,
rather than just the mean of this distribution. The results of this
calculation are shown in Figure \ref{fig:pdf_xbar} for $M_{\rm halo} =
10^{13} M_\odot$ and $M_{\rm min} = M_{\rm cool}$ at $z_{\rm on} =
6.66$. Note that the curves extend in some cases beyond $x_i = 1$,
which just implies that there are more than enough photons to ionize
the entire region in question. On small smoothing scales, there is
significant scatter in the ionization from host halo to host halo.
Indeed, there is some probability that a region will be {\em less}
ionized than the cosmic mean, although essentially all host halo
regions are more than $60-70\%$ ionized in this model.  This
scatter is potentially important for quasar proximity zone
observations.

In models where rarer sources dominate the ionizing photon budget, we
expect the tendency for overdense regions to reionize first to be
enhanced. Indeed, since reionization progresses most rapidly in such
models, they represent the most plausible scenarios in which the IGM
is considerably neutral at quasar turn-on yet highly ionized by $z=6$
(Figure \ref{fig:xbar_enhanced}).  On the other hand, we ignored
recombinations in our analysis, and these should lessen the tendency
somewhat for overdense regions to ionize first (e.g. Wyithe \& Loeb
2006b), although the precise impact of this process will depend on the
number of recombinations per hydrogen atom during reionization, which
is uncertain.

\subsection{Quasar Host Galaxy: Ionizing Photon Budget from Stellar and Quasar Activity}
\label{sec:star_qso}

A separate but related question regards the relative contribution of
starburst and quasar activity to the cumulative budget of ionizing
photons released by the {\em quasar host galaxy} alone (see also Yu \&
Lu 2005, Wyithe \& Loeb 2006a).  Indeed, the $z \sim 6$ quasar hosts
contain $\sim$ solar metallicity gas suggesting substantial levels of
prior star formation (e.g. Barth et al. 2003).  Our previous arguments
imply that neighboring galaxies reionize much of the surrounding IGM
before the highest redshift quasar turns on. Here, we focus solely on
the budget of ionizing photons produced within the quasar host galaxy
itself.

We can estimate the cumulative output of ionizing photons from quasar
and stellar sources on the basis of energetic arguments, combined with
assumptions regarding the radiative efficiency, the escape fraction of
ionizing photons, and the spectral energy distribution of emitted
radiation from quasars and stars respectively, along with an assumed
relation between black hole and stellar mass. A galaxy with a
stellar-mass, $M_\star$, produces a cumulative number of ionizing
photons (escaping into the IGM) given by $N_{\rm ion, \star} \sim
f_{\rm esc, \star} N_\gamma M_\star/m_p$. Here, $N_\gamma$ is the
number of ionizing photons produced per stellar baryon, $f_{\rm esc,
\star}$ is the escape fraction of stellar ionizing photons, and $m_p$
is the proton mass. The energy radiated while forming a black hole of
mass $m_{\rm BH}$ is $E_{\rm rad} \sim \epsilon_{\rm rad} m_{\rm BH}
c^2$, for a radiative efficiency of $\epsilon_{\rm rad}$.  Assuming
that a fraction $f_{\rm ion}$ of this energy comes out in ionizing
photons, with a fraction $f_{\rm esc, BH}$ of such photons escaping
into the IGM at a mean energy of $\langle h \nu \rangle_{\rm ion}$,
the cumulative number of ionizing photons produced by quasar activity
is $N_{\rm ion, BH} \sim f_{\rm ion} f_{\rm esc, BH} \epsilon_{\rm
rad} c^2/\langle h \nu \rangle_{\rm ion}$. We are interested in the
ratio of photons produced by quasar and stellar activity. We estimate
this ratio assuming that the local Magorrian relation connecting
stellar and black hole mass is upheld at high redshift, as supported
by the $z \sim 6$ black hole growth simulations of Li et
al. (2006). We assume typical values for the other parameters
(e.g. Cohn \& Chang 2006) arriving at:
\beqa  
\frac{N_{\rm ion, \star}}{N_{\rm ion, BH}} && \sim \frac{f_{\rm esc, \star}}{f_{\rm esc, BH}}N_\gamma \frac{M_\star}{M_{\rm BH}} \frac{\langle h \nu \rangle_{\rm ion}/f_{\rm ion}}{\epsilon_{\rm rad} m_p c^2} \\ \nonumber
&& \sim \frac{0.1}{1} \times 4000 \times 2000 \times \frac{30 eV/0.3}{10^8 eV} = 0.8.
\label{eq:ion_qstars}
\eeqa

For our fiducial parameters, the cumulative numbers of ionizing
photons from quasar and stellar activity associated with the quasar
host galaxy alone are already comparable, although uncertainties in
these parameters may allow this conclusion to change by $\sim$ an
order of magnitude. For example, if the quasar turns on before most of 
the stars and
lies off of the local Magorrian relation, the ratio of stellar to quasar 
photons will be smaller than in the above ratio. This late 
star formation scenario may be in 
tension with the presence of $\sim$ solar metallicity gas in the quasar host
as mentioned above. 
We view this simple estimate as a further argument that one needs to
consider the ionization from surrounding galaxies, in conjunction with
that from the quasar itself, in order to model high redshift
quasar proximity zones. On the other hand, note that although quasar
and stellar activity associated with the quasar host galaxy
cumulatively produce comparable numbers of ionizing photons, the
instantaneous photoionization rate from the quasar greatly exceeds
that from stars during the lifetime of the quasar (see \S
\ref{sec:1d_rt}).

%%%%%%%%%%%%%%%%%%%%%%%%%%%%%%%%%%%%%%%%%%%%%%%%%%%%%%%%%%%%%%%%%%%%%%%%%%%%%%%
\section{3D Radiative Transfer Calculations} \label{sec:3d_rt}
%%%%%%%%%%%%%%%%%%%%%%%%%%%%%%%%%%%%%%%%%%%%%%%%%%%%%%%%%%%%%%%%%%%%%%%%%%%%%%%

In this section, we use 3D radiative transfer calculations to
visualize the effects discussed above, and to detail the plausible
ionization state of the IGM when the highest redshift quasar turns
on. Here we aim to characterize the ionization field produced by high
redshift galaxies prior to quasar turn-on, and to quantify its
inhomogeneities.

\subsection{Simulations}
\label{sec:3dsim}

We use the 3D radiative transfer calculations from McQuinn et
al. (2006a). These simulations follow the growth of HII regions in a
cubic box of co-moving side-length $L_{\rm box} = 65.6$ Mpc/$h$.  The
radiative transfer calculations start from an N-body simulation run
with an enhanced version of Gadget-2 (Springel 2005), tracking
$1024^3$ dark matter particles, and resolving dark matter halos with
mass $M \gtrsim 2 \times 10^{9} M_\odot$. Ionizing sources are placed
in simulated dark matter halos, using a simple prescription to connect
ionizing luminosity and halo mass (Zahn et al. 2006).  Our fiducial
ionizing source prescription is identical to that described in Lidz et
al. (2006b): halos with mass larger than $M_{\rm min} = 2 \times 10^9
M_\odot$ host ionizing sources, with an ionizing luminosity
proportional to halo mass.   
The simulated dark matter density field is
then interpolated onto a $512^3$ Cartesian grid, and we subsequently
assume that the gas density closely tracks the simulated dark matter
density field (see Zahn et al. 2006 for a discussion). Finally,
radiative transfer is treated in a post-processing stage using the
code of McQuinn et al. (2006a), a refinement of the Sokasian et
al. (2001, 2003, 2004) code, which in turn uses the adaptive
ray-tracing scheme of Abel \& Wandelt (2002).

How sensitive are the HII bubble sizes at different stages of
reionization to the details of our modeling?  McQuinn et al. (2006a)
demonstrate that the size distribution of HII regions during
reionization depends most strongly on the ionized fraction, and is
less sensitive to the precise redshift at which a given ionization
fraction is reached, and other details of the reionization
process. These authors did find, however, larger HII regions at a
given ionized fraction if rare, highly biased sources produce most of
the ionizing photons. Furthermore, if mini-halos survive pre-heating
and are sufficiently abundant during reionization, this reduces the
characteristic HII region size (Furlanetto \& Oh 2005, McQuinn et
al. 2006a). Note that the minimum source mass in our fiducial simulation
is $\sim 10 M_{\rm cool}$, but even if this is a slight overestimate it
should have little impact on the bubble size distribution at fized
ionization fraction (McQuinn et al. 2006a).
In this paper we adopt the fiducial source prescription
mentioned above -- we refer the reader to the earlier papers for
details regarding the model dependence of our bubble size
distributions.

We now characterize the ionization field around plausible quasar host
halos. In practice, we examine outputs at several different redshifts
and ionization fractions.  However, based on the McQuinn et
al. (2006a) findings, we expect this to be roughly equivalent to
examining the ionization field at fixed redshift, yet varying
ionization fraction. Hence, we generally parameterize our partly
neutral models by the volume-weighted ionization fraction.

\subsection{Ionization Maps}
\label{sec:ion_map}

\begin{figure*}
\bc
\includegraphics[width=17.5cm]{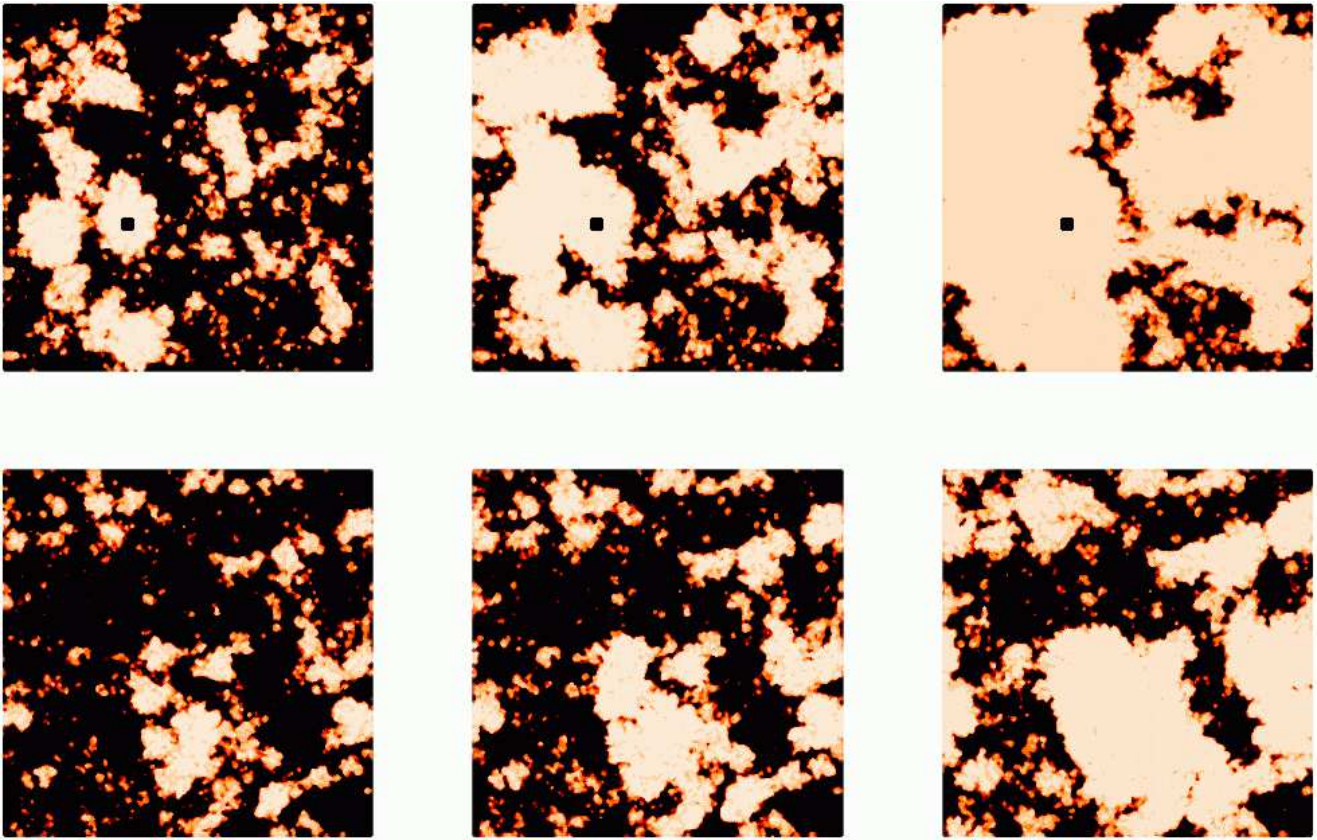}
\caption{Ionization fields around a plausible $z \gtrsim 6$ quasar host halo,
compared to the ionization in a typical region of the IGM.
{\em Top panel:} The ionization field in a thin slice ($0.13$ Mpc/$h$ thick, 
with a side-length of $L_{\rm box} = 65.6$ Mpc/$h$)
around a halo which may subsequently house a $z \gtrsim 6$ quasar.
The slices show different stages of reionization
in this overdense environment, when the globally-averaged, volume-weighted ionization
fraction is $\langle x_i \rangle = 0.31, 0.48$, and $0.70$ (from left to right). The white regions denote ionized gas, while black regions show neutral gas.
The black square indicates the location of the quasar host halo.
{\em Bottom panel:} For contrast, the ionization field in thin slices through
{\em random} regions of the IGM at the same global ionization fractions
as in the top panel.
Comparing the top and bottom panels, it is clear that reionization is
{\em accelerated} in the overdense regions that will subsequently 
host the $z \sim 6$ quasars. This results because galaxy formation
occurs earlier in overdense regions.
}
\label{fig:ion_overden}
\ec
\end{figure*}

First, we use our radiative transfer simulation simply to visualize the trends suggested by the 
analytic arguments of the previous section.  
In Figure \ref{fig:ion_overden} we show thin slices 
($1$ cell thick = $0.13$ Mpc/$h$) through the simulated ionization field at three different
stages during reionization (volume-weighted ionization 
fractions of $\langle x_i \rangle = 0.31, 0.48$, and $0.70$). In the left-hand side panels we
show the ionization field close to a plausible quasar host halo (with 
$M_{\rm halo} = 3 \times 10^{12} M_\odot$). For contrast, in the right-hand side panels
we show the ionization field through {\em random} slices of the same simulation outputs.

The figure provides a clear visualization of several points alluded to
in the previous section.  First, overdense environments are more
ionized than typical regions: the eventual quasar host halo is
surrounded by a large HII region blown out by galaxies born before the
quasar itself turns on. Second, the HII regions are themselves
correlated over large scales and so the host halo bubble tends to be
surrounded by large neighboring HII regions.  These features are seen
clearly at each of a few different stages during reionization by
contrasting the region containing the quasar host halo in the left
panel with the random regions shown in the right panel.  The bottom
two panels are likely more relevant than the top panels given the
arguments of the previous section which suggest $\langle x_i \rangle
\gtrsim 0.5$ at quasar turn-on.  Note also that our limited simulation
volume likely impacts our ability to capture large HII bubbles at the
end of reionization. In particular, while our general point should be
robust to our limited simulation volume, the detailed results may be
affected -- particularly at $\langle x_i \rangle = 0.7$ -- as one
might expect from the large HII regions shown in Figure
\ref{fig:ion_overden}.

\begin{figure}
\bc
\includegraphics[width=9.2cm]{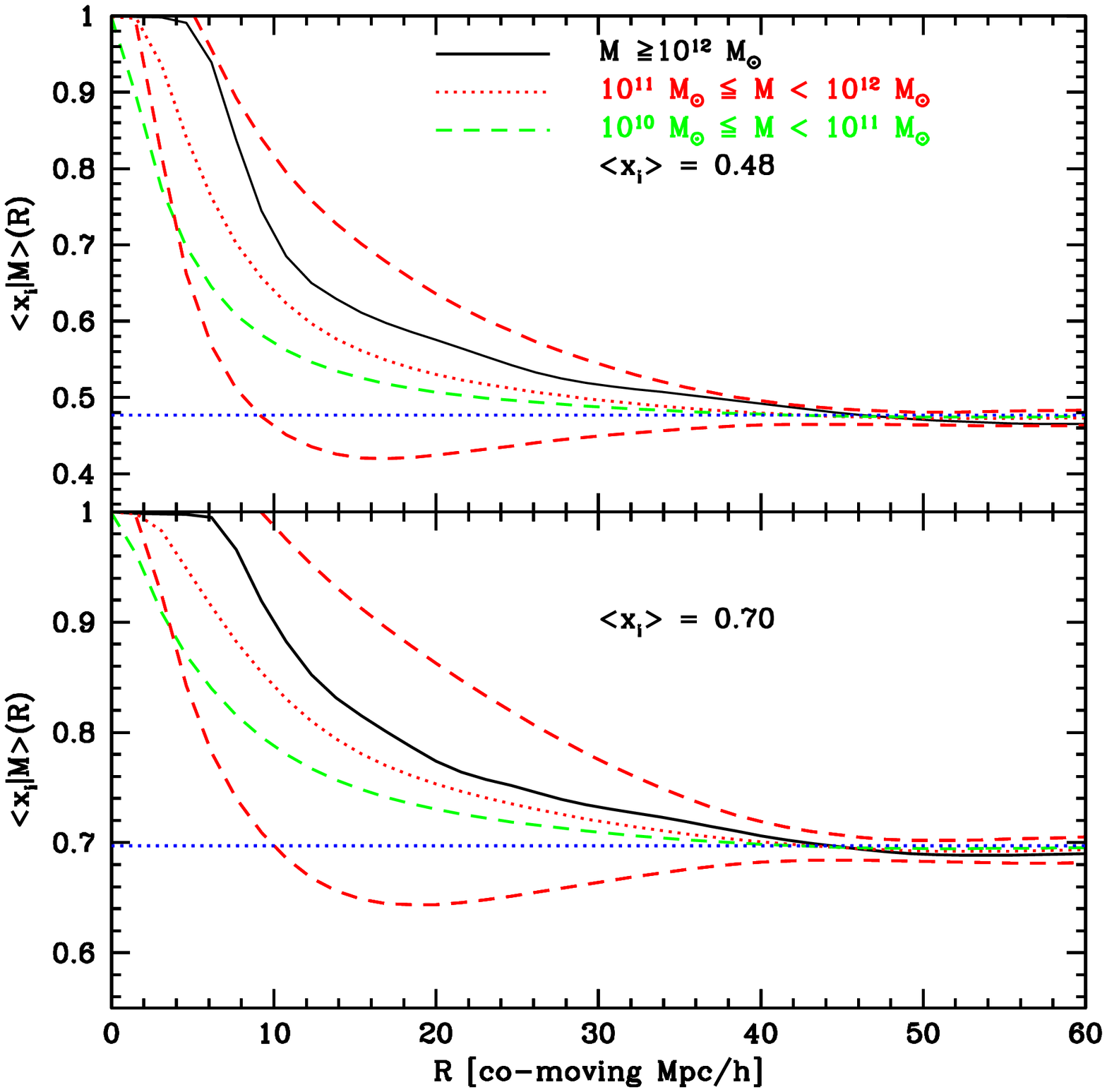}
\caption{Spherically-averaged ionization fraction centered on halos of different mass, as a function of
radius. The solid black, red dotted, and green dashed lines indicate the {\em average} ionization fraction
surrounding massive halos from the simulation, while the red dashed band gives the $1-\sigma$ spread across
different halos in the middle mass bin ($10^{11} M_\odot \leq M < 10^{12} M_\odot$). 
The horizontal blue dotted line indicates the ionization fraction averaged over the entire 
simulation volume. The {\em top panel} shows the average ionization around massive
halos when $\langle x_i \rangle = 0.48$, while the {\em bottom panel} gives
the same at $\langle x_i \rangle = 0.70$.
Reionization happens earlier around the overdense regions that house massive halos in our simulation, and hence the 
surrounding IGM is more ionized than typical regions. 
}
\label{fig:xbar_bias}
\ec
\end{figure}

\subsection{Statistical Description of Ionization Fields around Quasar Progenitors}

We can quantify the qualitative features of Figure
\ref{fig:ion_overden} in two useful ways. The first such statistical
measure is to compare the average ionization fraction in spheres of
different size, with each sphere centered on a massive halo. This
calculation is essentially a simulation version of the analytic
calculation shown in Figure \ref{fig:pdf_xbar}.  For comparison, we
show the results at each of $\langle x_i \rangle = 0.48$ (roughly the
most neutral case plausible when the $z \gtrsim 6$ quasars turn on, as
discussed in the previous section), and $\langle x_i \rangle =0.70$
for three different mass bins.  For each mass bin, regions
surrounding massive halos are on average more ionized than the cosmic
mean ionization level. As we discussed previously, this
`bias' results because overdense regions contain more halos and hence
ionizing sources and reionize earlier than typical regions.  This
bias is naturally most significant for the largest simulated
halos, which are likely hosts for the $z \sim 6$ SDSS quasars. Indeed,
Figure \ref{fig:xbar_bias} demonstrates that the {\em spherically
averaged} ionization around simulated halos with $M \gtrsim 10^{12}
M_\odot$ exceeds the cosmic mean ionization out to scales of $R \sim
40$ Mpc/$h$, and $75\%$ of the volume is typically ionized within
spherical shells of radii $R \lesssim 10$ Mpc/$h$, and $R \lesssim 25$
Mpc/$h$ at $\langle x_i \rangle = 0.48$ and $\langle x_i \rangle =
0.70$ respectively.  The true bias is potentially even larger, since
the $z \sim 6$ quasars may reside in even rarer, more massive halos
than contained in our simulation volume (Haiman \& Loeb 2001, Fan et al. 2001, Li et
al. 2006). Our most massive simulated halo at this redshift has a dark
matter mass of $M_h = 3 \times 10^{12} M_\odot$.

Also relevant is the level of halo-to-halo scatter in the spherically averaged ionization fraction.
In Figure \ref{fig:xbar_bias} we show the $1-\sigma$ scatter across halos in the mass 
bin with $10^{11} M_\odot \leq M < 10^{12} M_\odot$. The scatter is quite significant: for example,
there are regions of $R \gtrsim 10$ Mpc/$h$ centered on massive halos that are {\em less} ionized than the
cosmic mean, even though such regions are on average {\em more} ionized than the 
cosmic mean level. This scatter indicates that variations in the initial ionization field around
quasar host halos are substantial, and can potentially lead to significant sightline-to-sightline 
scatter in the proximity zone size -- provided that these observations indeed probe the pre-reionization IGM.
The simulation results shown here are also roughly consistent with the analytic calculations of Figures
\ref{fig:xbar_enhanced} and \ref{fig:pdf_xbar}.

\begin{figure}
\bc
\includegraphics[width=9.2cm]{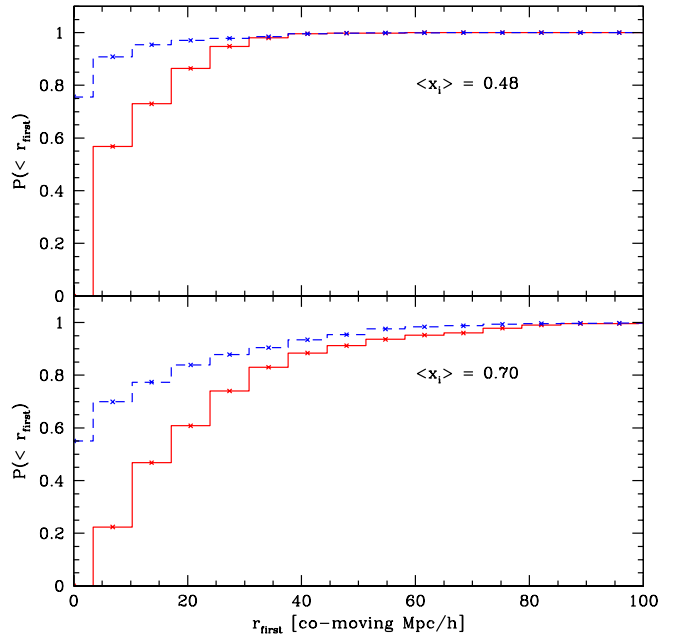}
\caption{Cumulative probability distribution for first crossing neutral cells.
The {\em top panel} shows results for an ionization fraction
of $\langle x_i \rangle = 0.48$, while the bottom panel shows the probability
distribution at $\langle x_i \rangle = 0.70$.
In each panel, the red solid histogram gives the first-crossing distance for random sightlines 
emanating from probable quasar host halos, before the quasar turns on. The 
blue dashed histogram shows,
for contrast, the first crossing distance for sightlines emanating from a {\em random} point in the IGM.
The first crossing distributions
are quite broad, and lines of sight from probable quasar host halos
may extend for a length comparable to the purported proximity
zone size ($L \sim 40 Mpc/h$) before touching {\em any} neutral cells, even prior to quasar turn-on. 
}
\label{fig:first_cross}
\ec
\end{figure}

While the spherically averaged ionization fractions shown above are
illustrative, a $1D$ statistic is more relevant for interpreting
quasar absorption spectra. As an example, we extend lines of sight
from the massive halos in our simulation box and calculate the distance traveled along
each line of sight before crossing neutral material.  In
practice we define a cell to be neutral when the HI photoionization
rate averaged over the cell is less than $10^{-5}$ times the cosmic
mean photoionization rate. Our general point is, however, insensitive
to this somewhat arbitrary definition.  The results of this
calculation are shown in Figure \ref{fig:first_cross}, where we
construct the cumulative probability distribution of `first neutral
crossings' calculated from 100 lines of sight, cast at random
directions from each simulated halo with $M_{\rm halo} > 10^{12}
M_\odot$. We consider the first crossing distribution when the
volume-weighted ionization fraction of the IGM is each of $\langle x_i
\rangle = 0.48$ and $\langle x_i \rangle = 0.70$.  The probability
distributions shown in the figure are quite broad, and the mean first
crossing distance from lines of sight towards quasar host halos is
large.  Specifically, at $\langle x_i \rangle = 0.48$, the mean and
$95 \%$ upper limit for the first crossing distance are $13$ and $28$
Mpc/$h$ respectively, while the corresponding numbers are $24$ and
$61$ Mpc/$h$ at $\langle x_i \rangle = 0.70$. Hence, even in partly
neutral scenarios, long skewers towards quasar hosts will traverse
entirely through ionized bubbles before the quasars themselves turn
on.  Indeed if the $z \gtrsim 6$ quasars turn on during the later
stages of reionization, a sizable fraction of skewers should pass
entirely through ionized bubbles out to length scales even larger than
$\sim 40$ Mpc/$h$ -- comparable to the purported size of proximity
zones inferred from the highest redshift quasar spectra
(e.g. Wyithe et al. 2005a). This indicates that the proximity zone
sizes deduced from quasar spectra are {\em not} primarily determined
by the volume-weighted neutral fraction of the IGM, as we discuss
subsequently.

In summary, if the $z \gtrsim 6$ quasars truly probe the
pre-reionization IGM, the initial ionization field prior to quasar
turn on will be inhomogeneous even on quite large scales. Moreover,
quasars are biased tracers, residing preferentially in overdense
regions, likely ionized before typical regions in the IGM. On
spherical average, this bias will be relatively small once one
averages over scales of $\gtrsim 30$ Mpc/$h$. Nonetheless, 1D skewers
through the IGM may pass entirely through ionized bubbles out to
considerably larger scales. This is another example of aliasing, where
fluctuations in a random field along skewers of a given length are
much larger than those from averaging the same field over spheres of
comparable diameter (Kaiser \& Peacock 1991).  These results
contradict the previous wisdom that $z \sim 6$ quasar proximity zones
probe sufficiently large scales that ionization inhomogeneities
average out, allowing one to model the surrounding IGM with some
uniform, yet low level ionization field.  In fact, our results suggest
that these large scale inhomogeneities significantly complicate the
interpretation of $z \sim 6$ quasar proximity zones.  In order to
understand the precise implications of these initial inhomogeneities,
we now consider the subsequent propagation of quasar ionization
fronts.

%%%%%%%%%%%%%%%%%%%%%%%%%%%%%%%%%%%%%%%%%%%%%%%%%%%%%%%%%%%%%%%%%%%%%%%%%%%%%%%
\section{Quasar Ionization Fronts and 1D Radiative Transfer} \label{sec:1d_rt}
%%%%%%%%%%%%%%%%%%%%%%%%%%%%%%%%%%%%%%%%%%%%%%%%%%%%%%%%%%%%%%%%%%%%%%%%%%%%%%%

In this section we follow the propagation of quasar ionization fronts using the calculations of
the previous section as `initial conditions' and considering, in particular, the difference
between the patchy-reionization models discussed above and partly neutral models with some
low level {\em uniform} ionization, as assumed by most previous modelers.

It is important
to keep in mind that our ultimate goal is to construct mock Ly$\alpha$ forest absorption spectra
for each scenario and compare with observations. In this sense, our results will be sensitive to the
precise values of the residual neutral fraction within the ionized bubbles shown in Figure
\ref{fig:ion_overden}. For example, if residual neutral fractions are at the $\sim 10^{-3}$ level, 
the Ly$\alpha$ absorption spectra will be very different than if residual neutral fractions 
are instead only $\sim 10^{-4}$.
This is in contrast to simulations of the 21 cm signal during reionization,
for which the bubbles of Figure \ref{fig:ion_overden} are simply `holes' of effectively zero signal, and
one is hence insensitive to the small residual neutral fraction in the interior of the bubbles (apart
from occasional dense clumps of more neutral gas which fill only a small fraction of the interior 
volume).
Simulating Ly$\alpha$ absorption spectra in the pre-reionization epoch is therefore
more challenging than
characterizing the large-scale 21 cm signal. Moreover, our 3D radiative transfer code assumes a sharp
ionization front, simply tracking the position of the edge of the ionization front, and assuming
ionization equilibrium within the front. Our 3D calculations also ignore helium and assume a 
uniform temperature within the front. These approximations speed up our 3D calculations substantially,
and they are likely very accurate for capturing the size distribution of HII regions during
reionization, and the 21 cm signal. However, they may be inadequate for our present purposes: here
we want the
precise residual neutral fraction and thermal structure within bubble interiors after ionization
from quasars. Our usual approximations are especially dubious given that quasars have a hard spectrum of ionizing 
photons.

In order to do optimize our method, we supplement our 3D analysis from
the previous section with more detailed 1D radiative transfer
calculations, tracking the subsequent ionization from the quasar.
Here our technique is similar to that of Bolton \& Haehnelt
(2006).  This approach is also natural for this application since
quasar absorption spectra probe only 1D skewers through the IGM.  Our
1D calculations self-consistently solve for the optical depth, the
ionization balance, and the temperature of the gas after the quasar
turns on. We do not track the hydrodynamic response of gas to the
ionization fronts. Of course our 1D approach is also an approximation, but we
expect it to be a good one.

\subsection{Model Parameters}
\label{sec:model_param}

Before proceeding with our 1D calculations, we briefly discuss several
uncertain parameters that go into our subsequent modeling.  The first
relevant quantity is the size distribution of HII regions at different
stages of reionization, as mentioned in the previous section. This
depends primarily on the nature of the ionizing sources and the
abundance of mini-halos and Lyman-limit systems (Furlanetto \& Oh
2005, Furlanetto et al. 2006c, McQuinn et al. 2006a).  In the present
paper, we consider only the fiducial ionizing source prescription
described in the previous section. If mini-halos/Lyman-limit systems
are abundant, then the pre-quasar HII regions will be smaller than
assumed here, while if rarer sources produce most of the ionizing
photons, the HII regions will be larger (see McQuinn et al. 2006a for
quantitative details).

\subsubsection{Hydrogen Photoionization Rates}
\label{sec:prates}

Even given the size distribution of HII regions, the precise hydrogen
photoionization rate within bubble interiors is somewhat uncertain. It depends
on the precise nature of the ionizing sources, the abundance of Lyman-limit
systems and the mean free path for ionizing photons in the bubble interiors,
and other factors.
In principle, one would
hope to model all of these quantities self-consistently. In practice, we believe that
the precise photoionization rates calculated from our 3D simulations for a given model 
are less robust than our predictions of the bubble size distribution and 21 cm signal.
Owing to this, we will extract only the spatial fluctuations in $\Gamma_{\rm HI}$ 
($\Delta_\Gamma(\vec{x}) = \Gamma_{\rm HI} (\vec{x})/\avg{\Gamma_{\rm HI}}$) directly
from our 3D simulation, and treat the spatial average as a free parameter. In our
fiducial calculations,  we somewhat arbitrarily 
normalize the volume-averaged 
HI photoionization rate to $\avg{\Gamma_{\rm HI}} = 10^{-13} s^{-1}$ (we adopt this value for
both our $\avg{x_i} = 0.48$ and our $\avg{x_i} = 0.7$ model). 
This is, however, similar to
the values measured directly from our simulation: $\langle \Gamma_{\rm HI} \rangle 
= 9.6 \times 10^{-14}
s^{-1}$ at $\langle x_i \rangle = 0.48$, and $\langle \Gamma_{\rm HI} \rangle = 2.2 \times 10^{-13}
s^{-1}$ at $\langle x_i \rangle = 0.7$. In \S \ref{sec:model_dep} we examine the sensitivity of 
mock Ly$\alpha$ absorption spectra to this choice.

For clarity, we pause here to point out the significant distinction between the hydrogen 
photoionization rate in a uniformly ionized medium, and that in a `patchy reionization'
model. In a uniformly-ionized IGM, photoionization equilibrium implies a one-to-one
relation between the neutral fraction at a given density and the hydrogen photoionization
rate. It is easy to see that the volume-averaged hydrogen photoionization rate in our
patchy reionization models will be vastly different than that in a uniformly-ionized
IGM of the same (low-level) mean ionization fraction.

Let us illustrate this point explicitly by considering two toy models.
In the first case,
representing patchy reionization, we imagine equal-sized ionized bubbles each with an interior
neutral fraction $X_{\rm IN} << 1$, filling a fraction $f$ (with $0 < f < 1$) 
of the volume of the IGM, which is
otherwise completely neutral. For simplicity, we neglect helium and consider an IGM with a 
uniform density and thermal state. In the second model, representing `uniform ionization',
the neutral fraction is identical at each location within the IGM with 
$X_{\rm HI} = \avg{X_{\rm HI}} = 1-f$. In each case, photoionization equilibrium tells us
that $\avg{\Gamma_{\rm HI}} = \avg{\alpha n_H} \avg{(1 - X_{\rm HI})^2/X_{\rm HI}}$. Now,
in a uniformly-ionized IGM we get 
$\avg{\Gamma_{\rm HI}}_{\rm uniform} = \avg{\alpha n_H} f^2/(1-f)$. In the toy patchy model,
we have $(1 - X_{\rm HI})^2/X_{\rm HI} \sim 1/X_{\rm IN}$ inside each ionized bubble, and 
$(1 -X_{\rm HI})^2/X_{\rm HI} \sim 0$ outside of ionized regions. Hence,
$\avg{\Gamma_{\rm HI}}_{\rm patchy} \sim \avg{\alpha n_H} f/X_{\rm IN}$. The ratio of the
volume-averaged photoionization rates is just 
$\avg{\Gamma_{\rm HI}}_{\rm patchy}/\avg{\Gamma_{\rm HI}}_{\rm uniform} 
\sim (1 -f)/(f X_{\rm IN})$. This will typically be a very large number: for example, if
$50\%$ of the volume is filled by ionized bubbles ($f \sim 1/2$) each with an interior neutral fraction
of
$X_{\rm IN} \sim 10^{-4}$, the volume averaged photoionization rate is a
factor of $\sim 10^4$ times larger in the patchy-reionization model than in the uniform model.
Hence the volume-averaged photoionization rates in our models are substantially larger
than a reader accustomed to a uniformly-ionized IGM might anticipate. There is also no
simple one-to-one relation between volume-averaged photoionization rate and neutral fraction,
although the photoionization rate should generally increase as a larger fraction of the IGM is
ionized, and each point in the IGM is influenced by more ionizing sources.

In order to give some sense for the model dependence of our adopted photoionization rates, we follow 
Furlanetto \& Oh (2005) and relate the hydrogen 
photoionization rate to the halo collapse fraction and mean free
path to ionizing photons. In this model, the (total) proper ionizing emissivity is given
by $\epsilon \sim \zeta \avg{n_H} df_{\rm coll}/dt$, where $\zeta$ is the ionizing efficiency parameter
of Equation (\ref{eq:xbar_delta}), $\avg{n_H}$ is the (average) proper abundance of hydrogen atoms, 
and $df_{\rm coll}/dt$
is the time derivative of the collapse fraction. Assuming that the mean free path to ionizing photons
scales as $\nu^{3/2}$ (Zuo \& Phinney 1993, \S \ref{sec:lls}), we then have:
\begin{eqnarray}
\avg{\Gamma_{\rm HI}} && \sim \zeta \avg{n_H} \frac{df_{\rm coll}}{dt} \sigma_{\rm HI} \frac{\alpha}{\alpha+1.5} 
\lambda_{\rm LL} = 10^{-13} s^{-1} \nonumber \\
&& \times \left[\frac{\zeta}{11}\right] \left[\frac{|df_{\rm coll}/dz|}{0.03}\right] 
\left[\frac{1+z}{7.5}\right]^{11/2} \left[\frac{\lambda_{\rm LL}(z)}{2 \rm{Mpc}}\right].
\label{eq:gammahi}
\end{eqnarray}

In this equation, $\sigma_{\rm HI}$ is the hydrogen photoionization absorption cross section at
the hydrogen ionization edge, which is numerically
$\sigma_{\rm HI} = 6.3 \times 10^{-18}$ cm$^2$. The quantity $\lambda_{\rm LL}$ represents
the proper mean free path to ionizing photons at the Lyman limit frequency (see \S \ref{sec:lls} for
comments on plausible values). The values of $\zeta$ and $df_{\rm coll}/dz$ above are chosen for
$M_{\rm min} = 10 M_{\rm cool}$ and so that $\avg{x_i} \sim 1$ at $z \sim 6$, assuming the mass function
follows the Press-Schechter (1974) form. It is interesting to note that our fiducial hydrogen 
photoionization rates are a bit higher than the values Fan et al. (2006) infer from their mean
absorption measurements on the basis of the Miralda-Escud\'e et al. (2000) density pdf. Our values are, however, more compatible with constraints from Becker et al. (2006a) derived using a 
lognormal model for the 
Ly$\alpha$ opacity. We do caution against over-interpreting this, given the large uncertainties
in our modeling. The above expression is only meant to loosely illustrate the model dependence of
our adopted photoionization rates.

\subsubsection{Quasar Lifetimes and Lightcurves}
\label{sec:qlife}

Next, let us consider the quasar lifetime and ionizing luminosity. In our fiducial model,
we adopt $t_q = 3.6 \times 10^7$ years for the quasar lifetime. This is our closest time output
to a Salpeter time (Salpeter 1964), $t_s \sim 4.5 \times 10^7$
years, assuming a radiative efficiency of $\epsilon_{\rm rad} = 0.1$ and
Eddington-limited accretion. This timescale is comparable to observational estimates of quasar
lifetimes (e.g. Martini 2004) and also roughly to the lifetime derived from numerical simulations
of black hole growth (Springel et al. 2005, Hopkins et al. 2005).
(In fact, the simulations of black hole growth indicate that the quasar
lifetime depends on luminosity, but the Salpeter time is a reasonable 
measure of the timescale during which most of the black hole mass
is accumulated; see e.g., Hopkins et al. 2006; Hopkins, Richards \& Hernquist 2007.)
However, the quasar lifetime
is still uncertain observationally by $\sim 2$ orders of magnitude (Martini 2004).
Consequently, we 
also consider $t_q \sim t_s/10$ in which case it is generally  
easier to locate the edge of a quasar proximity zone observationally (\S \ref{sec:mock_spectra}). 
Note, however, that decreasing the quasar lifetime increases the challenge of growing 
supermassive black holes by $z \sim 6$ (Haiman \& Loeb 2001, Li et al. 2006). In our fiducial model, we adopt an ionizing
luminosity of $\dot{N} = 1.8 \times 10^{57}$ photons per second, corresponding roughly to the
mean ionizing luminosity of the Fan et al. (2006) $z \sim 6$ quasar sample, provided their spectra
match the Telfer et al. (2002) quasar composite spectrum. For reference, the brightest quasar
in the Fan et al. (2006) sample is roughly $\sim 2$ times brighter than this, while the faintest
quasar is $\sim 3$ times fainter. For simplicity we assume here that quasars radiate
steadily at this luminosity for their entire lifetime (but see \S\ \ref{sec:conclusions}).

Finally, the level of small scale structure in the IGM is highly uncertain and can affect our
results. We examine the impact of small scale structure on our mock absorption spectra in the next
section: here we present results without including any `subgrid-clumping'. 

\subsection{Results from Our 1D Calculations}
\label{sec:1d_results}

As input for our 1D radiative transfer calculations we extract $20$
random lines of sight towards each of the five most massive halos in
our simulation box at $z \sim 6.6$. These halos have masses in the
range $1.4 \times 10^{12} M_\odot < M_h < 3.1 \times 10^{12} M_\odot$.
We simulate quasars at $z_q = 6.42$, corresponding to the highest
redshift source in the Fan et al. (2006) sample.  We project each
sightline at a random angle with respect to our simulation box,
interpolating the density and peculiar velocity fields (needed to
construct mock absorption spectra) from our N-body simulation. Each
sightline wraps around our periodic simulation box and extends for
$130$ co-moving Mpc/$h$. Our 3D calculations are not `in the
light-cone' and so we ignore redshift evolution in the ionization
state of the gas across each sightline.
 
For convenience in performing our radiative transfer calculation, and
generating mock quasar spectra, this interpolation is performed using
a fine grid with $1500$ cells.  Moreover, we interpolate the HI
photoionization rate from our 3D calculations onto each line of sight.
This specifies a (fluctuating) galactic ionizing background which is
incorporated as we calculate the subsequent ionization from our quasar
source. We approximate the galactic background ionization as fixed
over the quasar lifetime, $t_q \sim t_s \sim 4 \times 10^7$ yrs. From
the simulated HI photoionization rate, we specify a fluctuating
background HeI photoionization rate (our 3D calculations ignore helium
and hence do not track HeI photoionization), assuming a galactic
ionizing spectrum that follows a $\nu^{-2}$ power-law near the HI/HeI
photoionization edges. We further assume that our galactic sources do
not contribute any photons capable of doubly ionizing helium. Finally,
we assume that cells ionized by galaxies in our 3D calculations are
uniformly heated to $T = 10^4 K$ (gas temperature is not tracked
in our 3D calculations).  Neutral cells are assumed to be initially at
the CMB temperature, but our results would be unchanged for any other
small temperature.

Our 1D radiative transfer calculations follow the C2-ray scheme of Mellema et al. (2006a), generalized
to include helium and to track the thermal state of the gas in the IGM. 
We refer the reader to the Mellema et al. (2006a) paper for a description, and we will present 
details of our implementation elsewhere, but a few pertinent comments are as follows.
We make use of the fact
that, along the line of sight in the frame of the observer, the quasar front obeys the
{\em non-relativistic} I-front equation, and so there is no need to impose finite speed of light
constraints (see Cen \& Haiman 2000, Bolton \& Haehnelt 2006 and references therein). In order to rapidly compute
photoionization and photoheating rates, we adopt approximate forms for the frequency dependence
of the opacity between the HeI and HeII edges and above the HeII photoionization edge. This allows
us to compute photoionization and photoheating rates from lookup tables after specifying just the
optical depth in each of three frequency bins (see Shapiro et al. 2004 for a similar scheme). 
We use the thermal and chemical rates from Hui \& Gnedin (1997). Our code has been tested against
the usual analytic solutions for I-front propagation in a homogeneous IGM.

\begin{figure}
\bc 
\includegraphics[width=9.2cm]{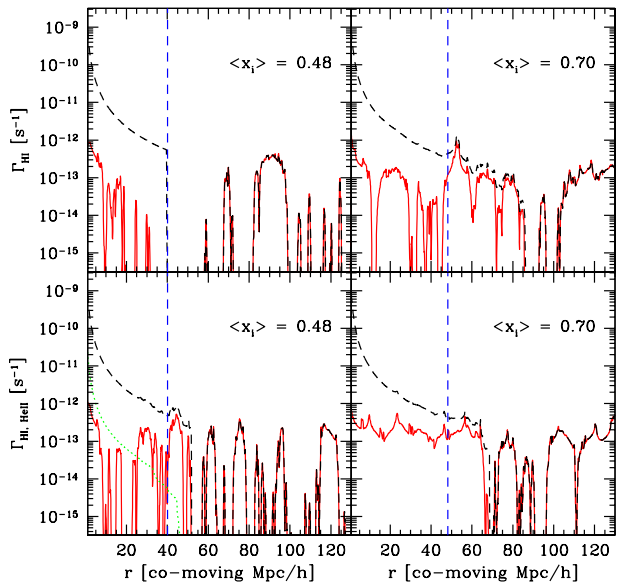}
\caption{Photoionization rate of HI and HeII from our 1D radiative
transfer calculations as a function of separation ($r$) from the quasar.
The {\em left panels} show the photoionization
rates along two random lines of sight through our simulation box, for
a case in which the quasar turns on when $48\%$ of the simulation
volume is ionized. The red solid line shows the initial photoionization rate contributed by galactic
sources, before the quasar itself turns on. The black dashed
line shows the photoionization rate after the quasar has shined for $t_q \sim t_s$, reflecting
the combined radiation field of the quasar and background galaxies. 
The blue dashed line shows the expected location of a quasar HII front in an IGM
with a {\em uniform} ionization level of $48\%$. The top line of sight intersects many
neutral patches, and the quasar front halts around the expected position in a uniform
medium. In the bottom panel, the front passes through longer stretches of ionized gas and extends
further. The green dotted line in the bottom panel shows the relatively broad quasar HeII front
(for visual clarity we omit the HeII curve in the other panels).
The {\em right panels} are identical for sightlines in which $70\%$ of the volume is ionized when
the quasar turns on. Since quasars are born into large HII regions, the typical sightline extends
further than expected in a uniform medium.
}
\label{fig:1drt}
\ec
\end{figure}

We show results from our 1D calculations in Figure \ref{fig:1drt} for
our fiducial model with initial (pre-quasar) volume-averaged
ionization fractions of $\avg{x_i} = 0.48$ and $0.70$.  Consider first
the combined HI photoionization rate from galaxies and quasars
(indicated by the black dashed lines in Figure \ref{fig:1drt}): at
small separations, the quasar strongly dominates the photoionization
rate and there is a relatively smooth $\sim 1/r^2$ fall-off, while the
HI photoionization rate fluctuates significantly at larger distances
from the quasar.  The extent of the $\sim 1/r^2$ run indicates the
maximum distance the quasar front reaches over the quasar lifetime
along a given sightline: the front halts at some distance upon
crossing sufficient quantities of neutral gas, and subsequently there
is only photoionization from background galaxies.  The photoionization
from background galaxies itself fluctuates owing to the presence of
neutral stretches of gas, since there is a {\em distribution} of
bubble sizes at each stage of reionization and the photoionization
rate is typically more intense in larger ionized bubbles, and because
the photoionization rate fluctuates from place to place within ionized
bubbles.

The red solid lines in Figure \ref{fig:1drt} indicate the initial
ionization provided by galaxies before the quasar itself turns on.
From the example sightlines, it can be seen that quasar ionization fronts
extend further along sightlines with larger levels of `pre-ionization'
from background galaxies. For example, the quasar front traverses more
neutral material, and extends less far, in the top-left panel than in
the bottom left panel, even though both adopt the
same quasar lifetime and volume-weighted ionization fraction. Another
interesting feature of Figure \ref{fig:1drt} is that the opacity to
ionizing photons at the Lyman-limit frequency from residual neutral
material amounts to $\tau_{\rm} \sim 1$ at $r \sim 50$ co-moving
Mpc/$h$. Hence, there is some fall-off in excess of $\sim 1/r^2$
for lines of sight in which the front extends to large distance,
although this is slightly disguised in our figure by upward
fluctuations in the galactic photoionization rate. In \S \ref{sec:lls}
we argue that our simulations may underestimate this residual opacity.
 
\begin{figure}
\bc
\includegraphics[width=9.2cm]{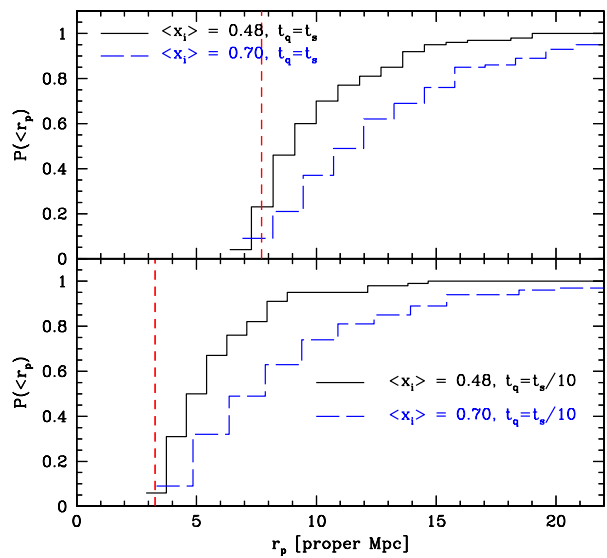}
\caption{Cumulative probability distribution of quasar ionization front
extents for partly neutral models. {\em Top panel:} The cumulative probability
distribution of quasar ionization front extents in models with initial
volume-weighted ionization fractions of $\avg{x_i} = 0.48$ and 
$\avg{x_i} = 0.7$. Here we assume a  
quasar luminosity of $\dot{N} = 1.8 \times 10^{57} s^{-1}$ and a quasar
lifetime of $t_q \sim t_s$.  The red dashed line indicates the expected quasar
front position in a uniform IGM with $\avg{x_i} = 0.48$ and $t_q \sim t_s$.
{\em Bottom panel:} The same for a quasar
lifetime of $t_q \sim t_s/10$. In each case, the histograms are quite
broad owing to the presence of large ionized bubbles created by galaxies
before quasar turn-on.
The fronts generally extend {\em further} than
in a homogeneous IGM of the same ionization fraction, since the quasars
tend to be born into large HII regions. Occasionally, fronts passing through
mostly neutral gas extend slightly less far than in the uniform case owing 
partly to overdense gas and helium absorption. The scalings for quasar front extent
expected in a uniform IGM,
$r_p \propto t_q^{1/3}/<x_{\rm HI}>^{1/3}$, are inaccurate owing to the
presence of large galaxy-created HII regions.
}
\label{fig:pdf_fronts}
\ec
\end{figure}

The example sightlines of Figure \ref{fig:1drt} illustrate 
a diversity of quasar front positions at a given $\avg{x_i}$, 
with a strong dependence
on the initial galaxy-generated ionization along each sightline, but
we require a statistical measure to characterize this quantitatively.
In Figure \ref{fig:pdf_fronts}, we show the cumulative pdf of ionization
front extent, averaged over our ensemble of $100$ sightlines for two
partly neutral models and two assumed quasar lifetimes. 
Here, and in what follows, we show results in units of proper Mpc to 
match the convention
of previous proximity-zone modelers.
The vertical
dashed lines in each panel illustrate the expected front position
in a uniform IGM with $\avg{x_i}=0.48$. In each case, we see the bias 
discussed in the previous
section: since quasars are born into large HII regions, the fronts extend
further on average compared to an IGM that is {\em uniformly} ionized
(with an ionization fraction equal to the volume-averaged ionization
fraction in the patchy reionization case). On the other hand, there
are also occasional sightlines that reach less far than expected in a uniform
IGM. These sightlines traverse longer neutral stretches than the
typical sightline. The presence of overdense gas, and absorption by
helium, also act to reduce the front extent compared to the expectation for
a uniform IGM. For the majority of sightlines, however, 
the fronts travel further than expected in a uniform IGM. The average
front extends $\sim 25\%$ ($\sim 30\%$) further in our patchy reionized
model at $\avg{x_i} = 0.48$ ($\avg{x_i} = 0.70$) -- assuming $t_q \sim t_s$ --
than in a uniform IGM.

Moreover, the sightline to sightline scatter in quasar front position
is quite large. For example, it takes until $r_p \sim 12.9$ ($18.1$)
proper Mpc to get to the $84\%$ ($97.5\%$) upper limit on quasar
front extent when $\avg{x_i} = 0.48$ and $t_s \sim t_q$. Clearly
quasar ionization fronts can extend much further than in a uniform
IGM, where one expects $r_p \sim 7.7$ Mpc for $\avg{x_i} = 0.48$, $t_s
\sim t_q$. The bias and scatter are still stronger when the quasar
lifetime is shorter, as illustrated in the bottom panel of Figure
\ref{fig:pdf_fronts}. This occurs because when the quasar lifetime is
short the quasar has less time to ionize the surrounding gas and the
front extent along an individual sightline is hence even more
sensitive to its initial pre-quasar ionization state. The histograms
also illustrate that the scalings expected in a uniform IGM, $r_p
\propto t_q^{1/3}/\avg{x_{\rm HI}}^{1/3}$, are inaccurate: the extent
of the quasar ionization fronts is sensitive to the presence of
galaxy-generated ionized pathways. This means that one needs a
significant data sample to infer the ionized fraction statistically,
as well as a model to connect the probability distribution of
proximity zone sizes with the volume-weighted ionization fraction.

It is also interesting to compare
the ionization front extents in these models with the `proximity' scale. This
scale is relevant for models in which the entire volume of the IGM is 
ionized and indicates where the photoionization rate from the
quasar drops to that of the background radiation field. Ignoring
residual opacity along the line of sight (this may overestimate the proximity
scale -- see \S \ref{sec:lls} for comments), this distance is 
$r_{\rm prox} \sim 18$ proper Mpc $\times$
$\left(\dot{N}/1.8 \times 10^{57} s^{-1}\right)^{1/2} 
\left(\Gamma_{\rm bcknd}/10^{-13} s^{-1}\right)^{-1/2}$. Notice that this scale
is comparable to the extent of quasar ionization fronts along many of the
sightlines in our partly neutral models, suggesting that it may be 
difficult to
distinguish models on the basis of their Ly-$\alpha$ forest proximity zones.
Another way to say this is as follows: for our fiducial parameters, even in partly
neutral models there is generally
little contrast between the photoionization rate at the edge of the quasar ionization
front and that provided by the background galaxies beyond the front 
(see Figure \ref{fig:1drt}). As we will see 
in the next section, 
this makes it challenging to locate the `edge' of the quasar front on the basis of 
Ly-$\alpha$ absorption spectra. The contrast between the photoionization rate at the
front edge and from the background galaxies will be larger than in our fiducial model if
$\avg{\Gamma_{\rm HI}}$ is smaller than we assume (background galaxies are less intense), or
if the quasar radiation field is more intense at the front edge.
The quasar will be more intense at the front
edge if the quasar lifetime is shorter (although even for a short lifetime quasar fronts
will extend to large distances along some sightlines, Figure \ref{fig:pdf_fronts}), and more
luminous. In practice, however, the brightest quasars in the Fan et al. (2006) sample are only
a factor of $\sim 2$ brighter than in our fiducial calculation.

Before constructing mock quasar spectra from our 1D calculations, we briefly discuss 
helium photoionization rates and the thermal state of the IGM (see also
Bolton \& Haehnelt 2006). 
The galaxies that reionize the IGM in our simulations have a soft
spectrum and so they mostly singly ionize helium, but do not doubly ionize it. Hence, before the quasar
turns on in our calculations, helium is mostly singly ionized and is doubly ionized only by the quasar 
itself. 
Because of this, the HeIII front lags behind the HII and HeII fronts. The HeIII front is, however, rather
broad owing to the long mean free paths for HeII ionizing photons. Both of these features are illustrated
by the green dotted line in the bottom left-hand panel of Figure \ref{fig:1drt}. The HeII front
(not shown in the figure for visual clarity), on the other hand, closely tracks the HII 
front in our calculation.
In principle, the position of the quasar HeIII
front provides interesting information regarding the ionization state of helium and the 
quasar lifetime: we can be fairly confident that HeIII, unlike HII, comes from
photoionization by the quasar itself. However,
HeII Ly$\alpha$ forest observations must be done in the ultraviolet band and are hence 
challenging (e.g. Shull 2004). Finally, initially neutral regions are raised to temperatures 
of $30-40,000$ K
by the photoheating of HI and HeII (see Abel \& Haehnelt 1999, Bolton \& Haehnelt 2006).
This quasar-heated gas recombines less rapidly, and is more thermally-broadened than the cooler
galaxy-ionized gas (which we assume to be initially uniformly heated to $10,000$ K.)

%%%%%%%%%%%%%%%%%%%%%%%%%%%%%%%%%%%%%%%%%%%%%%%%%%%%%%%%%%%%%%%%%%%%%%%%%%%%%%%
\section{Mock Quasar Spectra} \label{sec:mock_spectra}
%%%%%%%%%%%%%%%%%%%%%%%%%%%%%%%%%%%%%%%%%%%%%%%%%%%%%%%%%%%%%%%%%%%%%%%%%%%%%%%

At this point, the reader may be curious about two aspects of our line of reasoning.
The first point is that, starting with Wyithe \& Loeb (2004), several authors have
pointed out that the observed $z \gtrsim 6$ proximity zones are {\em small} and used this 
fact to argue for a partly neutral IGM. In the previous sections, we argued that quasar
ionization fronts may extend {\em further} than assumed by these authors in a partly
neutral IGM.  One might imagine that this {\em strengthens} the
case for a partly neutral IGM, or even places our results in conflict with observations.
Second, at least some of the sightlines in Figure \ref{fig:1drt} have
sharp ionization fronts and one might assume that it would be 
relatively easy to detect these sharp features in quasar absorption spectra.

In this section we will address both of these points. First, we illustrate that
in our fiducial model, the quasar is dim enough at the ionization front edge that
the average transmission through the forest is already very low (\S \ref{sec:average_trans}). Generally 
this means that
previous measurements of quasar ionization front extents are {\em underestimates}, as
pointed out previously by Bolton \& Haehnelt (2006) and Maselli et al. (2006). 
Mesinger \& Haiman (2004) also comment on
the difficulty of recovering quasar front positions from Ly$\alpha$ forest
spectra. On the other hand, they argue that the Ly$\beta$ region can be used
in conjunction with the Ly$\alpha$ forest to aid in recovering the position
of ionization fronts, although here one needs to contend with the foreground
Ly$\alpha$ forest. 
Here,
we describe and test (on mock spectra) an improved algorithm for determining the 
extent of quasar ionization fronts from Ly$\alpha$ forest spectra (\S \ref{sec:qfront_algorithm}). We then emphasize
the strong dependence of our results on the (highly uncertain) 
level of small-scale structure in the IGM, and other unknowns (\S \ref{sec:model_dep}-\S \ref{sec:lls}).

\subsection{Average Transmission in the Quasar Proximity Zone}
\label{sec:average_trans}

We construct mock quasar spectra from the neutral hydrogen density, temperature, and peculiar velocity fields
generated from our cosmological simulation, and 1D radiative transfer calculations,
in the usual manner (see e.g. Hernquist et al. 1996,
Hui et al. 1997). Note that we calculate the optical depth by convolving with
the full line profile, including the effects of both thermal broadening and the natural line width, since 
damping-wing absorption can be significant in our partly neutral scenarios (Miralda-Escud\'e 1998).
Since the damping wings are quite broad, we extend our sightlines slightly (ignoring redshift evolution in
the ionization state) in order to accurately calculate the optical depth near the edge of each sightline.

\begin{figure}
\bc
\includegraphics[width=9.2cm]{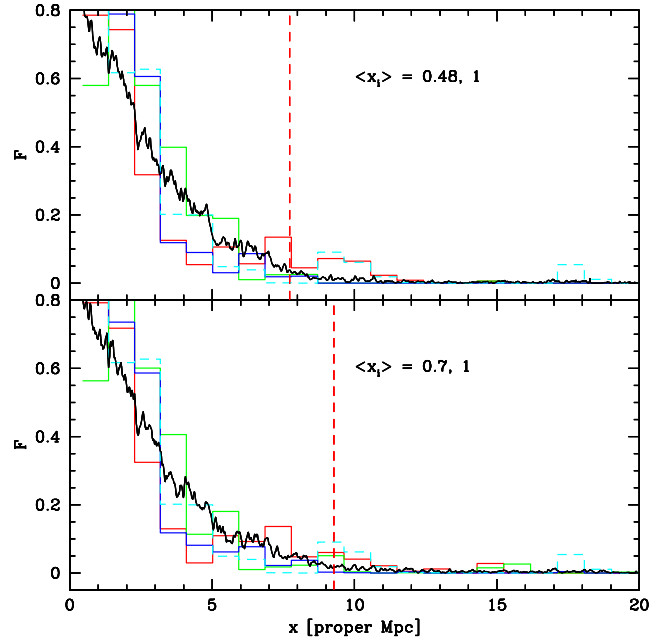}
\caption{Mock quasar 
spectra drawn from each of two partly neutral models and a completely ionized IGM. {\em Bottom panel:} The black solid line shows
the transmitted flux as a function of distance from the quasar, averaged over a large ensemble of 
quasar spectra for a patchy-reionization model with $\avg{x_i} = 0.7$. The red, green, and blue histograms
show the transmission along several example sightlines from the patchy model, smoothed as in 
Fan et al. (2006). The cyan dashed histogram shows an example sightline from a completely ionized IGM
($\avg{x_i}=1$) with the same mean hydrogen photoionization rate as our patchy models. The red dashed
line shows the expected ionization front extent in a uniform model with $\avg{x_i}=0.7$. In all
cases, there is transmitted flux beyond this scale, although in \S \ref{sec:model_dep} we show that this depends
on the amount of small-scale structure in the IGM. This transmission occurs because quasar fronts
tend to extend further in the patchy case than in the uniform, low-level ionization model, and because 
galaxy-created HII regions beyond the extent of the quasar front can transmit flux. It is hard to distinguish
between the highly-ionized and partly neutral models. {\em Top panel}: Identical to the bottom panel for a model
with $\avg{x_i}=0.48$.
}
\label{fig:flux_4los}
\ec
\end{figure}

First, let us consider the average transmission profile around several $z_q=6.42$ quasars hosts drawn from 
our simulation.
Specifically, we calculate the transmission using the 1D radiative transfer calculations of the previous section
along each of $100$ lines of sight, which were assembled from $20$ random lines of sight towards each of 
the five most massive halos in our simulation box. Clearly this is for illustrative purposes only, as present
data samples consist of only 3--4 quasars with $z \gtrsim 6.2$ (Fan et al. 2006).\footnote{One of the $4$ existing $z \gtrsim 6.2$ quasars
is a broad absorption line quasar, and has not been used in studies of the IGM (Fan et al. 2006, Mesinger \& Haiman 2006).}

To begin with, we consider our patchy reionization model with $\avg{x_i}=0.7$, and restrict our analysis to the blue
side of the quasar Ly$\alpha$ emission line. The $100$-sightline averaged transmission profile for this model 
is indicated by the solid black line in the top panel of 
Figure \ref{fig:flux_4los}. The sightline averaged transmission falls rapidly
with increasing distance from the quasar, a direct reflection of the decreasing quasar ionizing flux, which
falls off as $\Gamma_{\rm HI, qso} \propto 1/r^2$. Indeed, at $\sim 6$ Mpc proper from the quasar 
the {\em average}
transmission is only at the $10\%$ level. At still larger separations from the quasar there is only a small
residual flux, on average. The transmission at large separations results partly from sightlines that intersect
galaxy-created HII regions beyond the extent of the quasar HII front, and partly from sightlines where the
quasar HII front itself reaches large distances by passing through several galaxy-generated HII regions. 
The transmission extends significantly further than quasar fronts do in a uniform IGM
of the same ionization fraction, as illustrated by the vertical red dashed line in the figure (see \S 
\ref{sec:model_dep} for caveats). 
These features just illustrate the impact of pre-quasar HII regions and 
the extended and fluctuating photoionizing flux of Figure
\ref{fig:1drt} on the transmission in the Ly$\alpha$ forest. 

The {\em top panel} of Figure \ref{fig:flux_4los} is
very similar to the bottom one, except that in this panel the mock spectra are drawn from a model where reionization
is less advanced when the quasar turns on ($\avg{x_i}=0.48$). Comparing the solid line in the two panels,
one sees only very subtle differences in the average transmission between the two models.

Note that in our fiducial model, we do typically get some transmission through the Ly$\alpha$ forest
beyond the quasar range of influence
even when the IGM is $50\%$ neutral (Furlanetto et al. 2004b). This transmission may be overestimated
if the IGM is more clumpy than in our model, as we argue subsequently, or if we overestimate the
galactic photoionizing background. In order to get transmission through the IGM when it is half
neutral, there must also be sufficiently long stretches of ionized gas so that the damping wings
from neutral gas on each side of an ionized stretch do not overlap (Miralda-Escud\'e 1998). 
If we approximate the IGM as homogeneous on each side of an ionized stretch, with neutral fraction
$X_{\rm HI}$, the damping wing opacity (including both redward and blueward of the ionized stretch)
amounts to $\tau_{\rm dw}=1$ at the center of the ionized region of size
$L \sim 4 X_{\rm HI} \sigma_\alpha R_\alpha n_{\rm H}(z) (c/H(z))^2/\pi$ (Miralda-Escud\'e 1998). 
Here, $\sigma_\alpha$ is the Ly$\alpha$ cross-section, $R_\alpha = 2.02 \times 10^{-8}$ is related
to the natural line width, and the other symbols have their usual meanings. When the IGM is
$50\%$ neutral, this estimate -- which should be an overestimate close to the quasar -- 
requires a stretch of $L \sim 1.8$ proper Mpc.  Paths of this length are not rare in our
calculations when the IGM is $50\%$ neutral. Hence, resonant absorption may be more of an obstacle
for transmission in the partly neutral IGM, than damping wing absorption (Furlanetto et al. 2004b).

Let us now consider transmission spectra on a sightline-by-sightline basis. This is more observationally
relevant than the 100 sightline-averaged curve described above, given the limited observational samples available
presently. We follow Fan et al. (2006) and top-hat filter our mock spectra at $20 \Ang$ resolution, even though
their spectra have intrinsically higher resolution. This extra smoothing is performed because
as emphasized by Fan et al. (2006), the proximity scale, defined as the first time the transmission crosses
some threshold value, is naturally sensitive to the smoothing scale of the observations. Example lines of sight
are shown by the solid, colored histograms in Figure \ref{fig:flux_4los}. For comparison, we also show a sample
line of sight (cyan dashed histogram) drawn from a model in which the entire volume of the IGM is ionized.
One can immediately see that it is difficult to distinguish between different models for the volume-ionized fraction 
on the basis of a few absorption spectra alone. The main obstacle is that the transmission is generally very low
close to the edge of an ionization front in our models. Comparing with the highest redshift bin in 
Figure 14 of Fan et al. (2006) it seems, however,
that each of our models overproduces the transmission at large distances in comparison to the observations.
We return to this shortly in \S \ref{sec:model_dep}.

First, in order to further examine our ability to discriminate between models, we construct histograms of  
first-threshold
crossing distances from our smoothed spectra. This has been used by Fan et al. (2006) and others 
as a diagnostic for ionization-front position. Here we simply record the closest distance to the quasar 
at which the smoothed-transmission falls below a given threshold, $F_{\rm thresh}$. Using $F_{\rm thresh} = 0.1$ as in Fan
et al. (2006), we find little difference between models. This is because the IGM is opaque enough at these redshifts
that the transmission generally falls below this threshold even within the quasar ionization front, making this
measure a poor tracer of ionization front extent.
Using a smaller threshold is an improvement, but it is still very difficult to distinguish between models since even the
smoothed spectrum frequently falls below the threshold before reaching the front edge. If one chooses a still 
smaller
threshold, then one picks up transmission from background galaxies and misidentifies the front position. Smoothing each spectrum with a $20 \Ang$ filter also partly 
washes out distinctions between
models, since the transmission spikes are generally very narrow.

\subsection{Improved Algorithm for Determining Quasar-Front Extents}
\label{sec:qfront_algorithm}

Can we devise a better algorithm to locate the extent of quasar ionization fronts and distinguish between
different models?  Our basic task is to identify a scheme for locating an `edge' in a quasar absorption spectrum:
shortward of the edge, the quasar contributes to the ionizing flux, while longward of the edge only background
galaxies contribute. We propose a simple variant of the Fan et al. (2006) first threshold-crossing scheme, based on 
locating the {\em furthest} distance from a quasar,
beyond which the (un-smoothed) transmission crosses a given threshold. In locating the `last-crossing' distance, we consider only pixels within
$20$ proper Mpc of the quasar. We find that this `last-crossing' scheme
is a better diagnostic of quasar front position. It is also advantageous to use the full un-smoothed spectrum to 
preserve as much information as possible, although some smoothing will be necessary for sufficiently noisy data. 

\begin{figure}
\bc
\includegraphics[width=9.2cm]{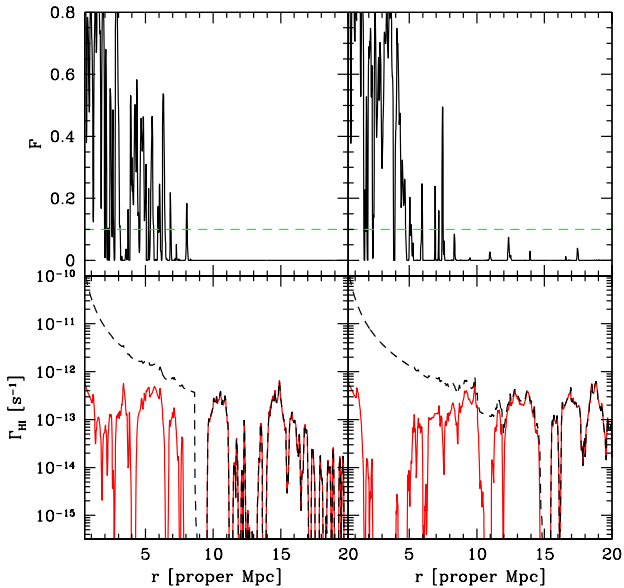}
\caption{Illustrative examples of our front-locating algorithm.
Here we show an example sightline where our method performs well, and an example where it performs
poorly. Each sightline is drawn from our model with $\avg{x_i} = 0.48$. {\em Left-hand panels:} Well-recovered front 
position example.
The ionization field ({\em bottom panel}), and (un-smoothed) transmission through the Ly$\alpha$ 
forest ({\em top panel}) for a sightline
where the edge of the quasar ionization front is recovered relatively accurately. As in Figure \ref{fig:1drt}
the red solid line shows the pre-quasar photoionizing field, while the black dashed line shows the combined
quasar and galaxy field. The green dashed line in the top panel indicates an example flux threshold
used in recovering the front position. The last place where the transmission crosses the flux threshold is close
to the true front position, which corresponds to where the photoionization rate first drops off in the bottom panel
(near $\sim 8-9$ proper Mpc). Along this sightline, in spite of the significant galactic photoionizing background,
there is no gas that is sufficiently underdense to transmit flux at greater separations from the quasar. 
{\em Right-hand panels}: Poor-recovery example. In this case, the quasar front extends to a considerably
larger distance ($r \sim 15$ proper Mpc). The IGM, however, is opaque enough to prevent transmission in excess
of the flux threshold beyond $r \sim 8$ proper Mpc. Consequently, along this sightline the true front position
is underestimated by nearly a factor of $\sim 2$.
}
\label{fig:front_recovery}
\ec
\end{figure}

The basic idea of our last-crossing threshold scheme is illustrated by the example sightlines shown 
in Figure \ref{fig:front_recovery}. A clean case is shown by the sightline in the {\em left-hand} panels,
drawn from our fiducial model with $\avg{x_i} = 0.48$. Here, the transmission fluctuates and is reasonably 
substantial before declining beyond $r \sim 8-9$ proper Mpc. In this case, the last place along the
line of sight where the transmission crosses a threshold of $F_{\rm thresh} = 0.1$ is a reasonable
indicator of quasar ionization front extent, as can be seen in the figure. Along this sightline, it is
a slight underestimate (by $\sim 10\%$), owing both to damping wing absorption and since one requires
a sufficiently underdense region to allow transmission through the forest. (Note that peculiar 
velocities also impact the mapping between observed and actual position.)
Note that there is also a substantial
galactic component to the ionizing flux at larger distances along this sightline. In this case, the surrounding
gas is, however, too dense to transmit flux through the Ly$\alpha$ forest.

Along other sightlines, it can be quite difficult to locate the position of the quasar ionization front, and
even our improved algorithm can lead to inaccurate results. This is illustrated by the example sightline in
the {\em right-hand} panels of Figure \ref{fig:front_recovery}. Along this line of sight, the quasar ionization
front extends to a larger distance ($r \sim 15$ proper Mpc), but fails to produce a transmission spike -- i.e.,
transmission exceeding our flux threshold -- beyond
$r \sim 7-8$ proper Mpc. This is again the difficulty mentioned in the previous sub-section: the transmission
at the quasar ionization front edge is generally low, and there is no guarantee of surrounding gas that is underdense
enough to transmit any flux. If we lower the flux threshold slightly, our algorithm will pick up the more distant
quasar-created transmission spike. However, if we lower the threshold too much, our algorithm will begin to identify
the still more distant galaxy-generated spikes. 

\begin{figure}
\bc
\includegraphics[width=9.2cm]{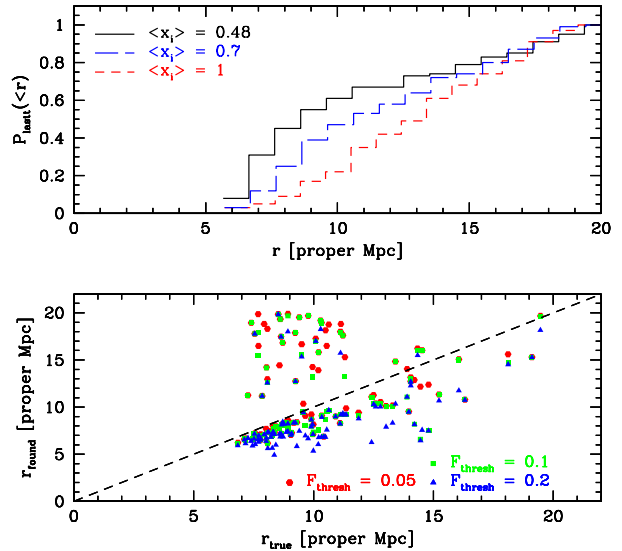}
\caption{Test of quasar front position recovery, and power to distinguish between models. 
{\em Bottom panel}: Scatter plot comparing the true quasar front position (x-axis) along a sightline, with the
front position
recovered using our last-crossing algorithm (y-axis). We show the recovered front positions for each of three flux thresholds.
The dashed line indicates perfect recovery: points underneath the line indicate sightlines where our algorithm
underestimates the true front position, while points above the line indicate overestimates. {\em Top panel}: The cumulative pdf of last-crossing positions (with $F_{\rm thresh}=0.1$)
for $100$ sightlines drawn from models with 
each of $\avg{x_i} = 0.48, 0.70$ and $1$ and all other parameters fixed at their fiducial values. This illustrates that
even if the error on the recovered front position along an individual sightline is large, we can still hope to 
distinguish
between models given a sufficiently large sample of sightlines. However, the details of this panel are subject
to several model uncertainties, particularly the level of small scale structure in the IGM (see \S
\ref{sec:model_dep}).
}
\label{fig:test_recovery}
\ec
\end{figure}

In order to characterize the accuracy of our algorithm more quantitatively, we locate the last threshold crossing
position for
each of $100$ sightlines drawn from our fiducial model at $\avg{x_i}=0.48$ and compare this with the true
quasar front position. The results of this test are shown in the bottom panel of Figure \ref{fig:test_recovery}.
Concentrating on the locus of points near $r \lesssim 8$ proper Mpc, we see that our front recovery algorithm
works reasonably well for small quasar front extents along many sightlines. 
In this case, the true front position is underestimated: this owes to damping wing absorption, and since one
needs to pass through an underdense region to transmit any flux. The recovered
front position hence, loosely speaking, indicates the last underdense region within the quasar front. The error in the recovered front
position can, however, be large as one can see from the many points lying above the dashed line in the figure that
indicates
perfect front recovery. These points correspond to sightlines where the galactic ionizing background beyond the quasar 
front
gives rise to transmission through the Ly$\alpha$ forest. Along these sightlines, the front-locating algorithm
is fooled into thinking that galaxy-created transmission spikes are associated with flux from the quasar, resulting in
an overestimate of quasar front position.
This source of error is reduced by increasing the flux threshold -- and hence reducing the occurrence of 
`false positives' -- but increasing the flux threshold in turn causes one to further underestimate the 
short-length quasar front positions. Note also that quasar ionization fronts really do extend to large distances
along some sightlines, so it is not viable to simply throw sightlines with large front extents out of the analysis.
Doing so could easily bias our constraints. Quantitatively, the true front position is recovered most accurately on
average for $F_{\rm thresh} = 0.05, 0.1$, in which case the true front position is identified to 
within $\pm 20\%$ along $\sim 40\%$ 
of sightlines for this model. There are however significant outliers as one can see clearly from the figure.

Even though the error in front positions from individual sightlines may be substantial, one can still hope to 
distinguish between different models statistically, albeit in a model-dependent manner. This is illustrated
by the top panel of Figure \ref{fig:test_recovery}. Here we show the cumulative pdf of recovered front positions
for $F_{\rm thresh} = 0.1$ at $\avg{x_i} = 0.48, 0.70$, and $1$, with our other model parameters fixed at their
fiducial values. In this example, we take the idealized case of $100$ quasar spectra with perfect signal-to-noise,
resolution and identical quasar luminosity and lifetime. One can clearly see the tendency towards larger
last threshold crossing distances as a greater fraction of the volume is ionized, as expected. In the context of 
this simplified calculation where, for example, all of our quasars have identical lifetimes and luminosities,
we expect $\sim 10$ $(30)$ spectra are needed to rule out a completely ionized model at 
$\sim 1-\sigma$ $(2-\sigma)$ if the real universe 
has $\avg{x_i} = 0.48$.\footnote{This significance
is estimated by drawing many random realizations of $N_{\rm spec}$ last crossing distances 
from the
$\avg{x_i} = 0.48$ model, which is our assumed `true' model. For each set of 
$N_{\rm spec}$ last crossing distances, we calculate their cumulative probability distribution
and the Kolmogorov-Smirnov probability that our $\avg{x_i} = 1$ model is drawn from the true
model. We estimate that $N_{\rm spec}$ spectra are required to distinguish the two models
at $1-\sigma$ $(2-\sigma)$ when $84\%$ $(97.5\%)$ of realizations have a Kolmogorov-Smirnov
probability of $16\%$ $(2.5\%)$.}

As we emphasize in the next subsection, there are important model uncertainties that could impact these
predictions. Specifically, the precise mapping between hydrogen photoionization rate and transmission is
sensitive to the level of small scale structure in the IGM. If there is less transmission at a given
hydrogen photoionization rate than in our model (as may be indicated by the data), then the distance beyond which
it becomes hard to locate the true front position ($r \sim 8-9$ proper Mpc in Figure \ref{fig:test_recovery}) will
shrink, and make it harder to distinguish models. On the other hand -- in comparison with our fiducial case -- 
increasing the quasar luminosity while
decreasing the quasar lifetime, and/or decreasing the intensity of the photoionizing background should make
it easier to locate ionization front edges. Finally, following Mesinger \& Haiman (2004), applying our algorithm
to the Ly$\beta$ forest may be helpful, although here one needs to deal with the foreground Ly$\alpha$ forest.
Furthermore, while our last-crossing algorithm is easy to implement, it is not `optimal' in any sense. One might
be able to improve on it by, for example, recording the length of dark gaps after last threshold crossing along
each sightline. Moreover, if one can indirectly extract the signature of damping wing 
absorption from the data (Mesinger \& Haiman 2004, 2006) this would be a smoking gun for
a partly neutral IGM, although uncertainties in the clumpiness of the IGM may complicate
this effort (\S \ref{sec:model_dep}).

\subsection{Sensitivity to Model Details}
\label{sec:model_dep}

At this point, we need to emphasize that the `mapping' between photoionization rate and transmitted flux is model 
dependent, and may not be well captured by our moderate resolution N-body simulations. 
Specifically, our transmission profiles depend on the clumpiness of the IGM and the mean photoionization rate
within galaxy-generated HII regions beyond the quasar range of influence. 

We begin by considering the clumpiness in the IGM. {\em After} reionization, the relevant scale is
the filtering scale (Gnedin \& Hui 1998), the distance over which baryonic fluctuations are smoothed owing
to finite gas pressure. According to Gnedin \& Hui (1998), the filtering scale can easily be a factor
of $\sim 10$ smaller than the linear Jeans scale close to reionization, since the gas takes some time
to respond to prior photoheating. In our calculations, we assume that the baryons directly trace the
dark matter, but density fluctuations are smoothed out by the finite resolution of our
simulation. The Nyquist frequency of our simulation grid is $k_{\rm Nyq} \sim 25 h$ Mpc$^{-1}$.
For comparison, if the gas in the IGM is isothermal with a 
temperature of $\sim 10^4 K$ at $z=6.5$, then the linear
Jeans wavenumber is $k_J \sim 15.7 h$ Mpc$^{-1}$. Given that the filtering wavenumber $k_F$ should be
at least several times -- and potentially more than an order of magnitude -- larger than the Jeans
wavenumber (Gnedin \& Hui 1998), we likely {\em underestimate} the small scale structure in 
the gas density field.
This is a consequence of our finite
simulation resolution ($k_{\rm Nyq} \sim 25 h$ Mpc$^{-1}$) and hence, even though
we assume that the baryonic density field directly traces that of the dark matter,
we may underestimate the clumpiness in the baryon distribution.

This could impact the results shown thus far in two ways. First, we underestimate
the number of recombinations during reionization and, consequently, we underestimate the
residual neutral fraction within previously ionized cells. Moreover, quasar ionization fronts
should extend slightly less far than they do in our calculations (see the next sub-section, \S \ref{sec:lls}, for
related discussions). Second, transmission in the
high redshift Ly-$\alpha$ forest is very sensitive to the abundance of rare voids in the IGM (e.g.
Oh \& Furlanetto 2005, Becker et al. 2006a). Our low simulation resolution artificially suppresses
the number of low-density excursions in the gas density field.

In order to explore the impact of the first deficiency in our modeling, we now incorporate a simple model
for sub-grid clumping into our calculations. We first estimate plausible levels for the clumpiness 
in the 
baryon distribution close
to reionization.  (For 
recent
related calculations see Kohler et al. 2005, Mellema et al. 2006b, 
McQuinn et al. 2006a, Trac \& Cen 2006; our present work is similar to that of McQuinn et al. 2006a.) 
We estimate a volume-averaged clumping factor
for the baryon distribution by taking $C_V = 1 + \langle \delta_b^2 \rangle$, with
$\langle \delta_b^2 \rangle \sim \int d^3k/(2 \pi^3) \ \rm{exp}(-2 k^2/k_F^2) P_{\rm dm}(k)$ 
and we
adopt the Peacock \& Dodds (1996) fitting formula for $P_{\rm dm}(k)$. 
This calculation may be an overestimate for two reasons. First, the relevant quantity for
our reionization calculations is the clumping factor of the {\em ionized} gas; our clumping
calculation should exclude the highly overdense interior regions of dense clumps which, in 
reality, self-shield and remain neutral. Second, recombinations within the massive
halos that host ionizing sources are -- to some extent -- incorporated in the escape fraction parameter
in our 3D calculations, and should not be double-counted.
Moreover, we should keep in mind that our assumed
relation between baryonic and dark matter fluctuations,  
$P_{\rm bar}(k) = \rm{exp}(-2 k^2/k_F^2) P_{\rm dm}(k)$, is approximately correct only 
according to linear
theory, and may not be accurate in the fully non-linear regime -- 
especially if reionization is not yet complete, and heating is highly inhomogeneous.

We contrast our semi-analytic clumping factor estimates
with the variance of our simulated density field smoothed on the scale of the simulation
mesh, $1 + \langle \delta_b^2 \rangle_{\rm mesh} =  1 + \langle \delta_{\rm dm}^2 \rangle_{\rm mesh} = 5.3$. 
(The first equality simply results because we assume that baryons precisely trace our simulated dark matter
density field.)
The precise value of the filtering
scale is quite sensitive to the details of the thermal history of the IGM near
reionization.
As a simple example, consider gas that is completely cool before some
$z_{\rm heat}$, at which point it is uniformly raised to a temperature $T_0$, subsequently
cooling as $T \propto 1/(1+z)$ (Gnedin \& Hui 1998). In this case, $k_F = 3.4 k_J$ if
$z_{\rm heat} = 10$, and $k_F = 9 k_J$ if $z_{\rm heat} = 7.5$. If the gas is then isothermal
at $10^4 K$ at $z \sim 6.5$ so that $k_J = 15.7 h$ Mpc$^{-1}$, our clumping factor 
estimates are $C_V = 10.2$ and $C_V = 22.8$ for $k_F = 3.4 k_J$ and $k_F = 9 k_J$ respectively.
These values are comparable to results from the numerical simulations of 
Mellema et al. (2006b) and Trac \& Cen (2006), but are considerably higher than the 
fluctuation level on our simulation mesh, $1 + \langle \delta_b^2 \rangle_{\rm mesh} = 5.3$.

We incorporate this extra small scale structure in our calculations by adopting a sub-grid clumping factor. We
consider two models for sub-grid clumping.
In each case
we adopt a uniform sub-grid clumping factor, ignoring any density dependence and any
cell-to-cell scatter. Our sub-grid models result in clumping
factors of $C_V = 10$ and $C_V=20$, as motivated above. We include the sub-grid clumping models only
in our $1D$ radiative transfer step, and not in the 3D calculations, since 
McQuinn et al. (2006a) found that similar clumping prescriptions
do not significantly impact the size distribution 
of HII regions during reionization at a given ionization fraction.
Incorporating sub-grid clumping boosts
the residual neutral hydrogen abundance within previously 
ionized cells by a factor of $\sim C_{\rm sub-grid}$,
and results in a modest reduction in the extent of quasar ionization fronts. 

\begin{figure}
\bc
\includegraphics[width=9.2cm]{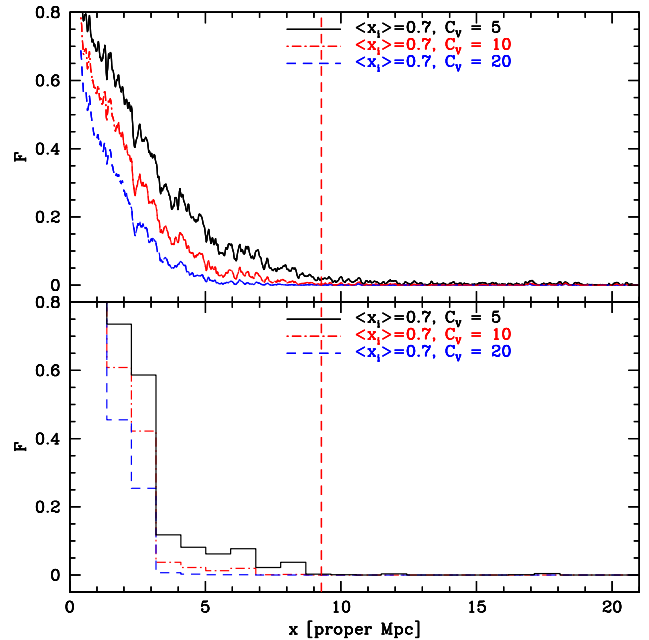}
\caption{Sensitivity of mock spectra to clumping in the IGM. {\em Top panel:} 
The solid lines show the transmission averaged over an ensemble of $100$ mock spectra
in models with volume-averaged clumping factors of $C_V = 5$ (i.e., no sub-grid
clumping), $10$ and $20$ (from top-to-bottom, black, red, and blue lines respectively).
The vertical red dashed line is the same as in previous plots.
{\em Bottom panel:} The histograms show a single example spectrum, smoothed as in Fan
et al. (2005), drawn from each of the above models. The mock spectra are more
sensitive to the level of clumpiness than to the volume-averaged neutral
fraction in the IGM (see Figure \ref{fig:flux_4los} to see the sensitivity to 
volume-averaged ionization fraction). Indeed, a model with $C_V = 10$ and a uniform
ionizing background ($\avg{x_i} = 1$) with the same mean intensity 
($\avg{\Gamma_{\rm HI}}=10^{-13} s^{-1}$) yields a nearly identical ensemble-averaged
transmission curve to our corresponding $\avg{x_i} = 0.7$ model.  
}
\label{fig:mock_v_clump}
\ec
\end{figure}

We show mock spectra drawn from our models with sub-grid clumping in Figure
\ref{fig:mock_v_clump}.  Owing to the increased residual neutral fraction within ionized
cells, increasing the amount of small scale structure significantly reduces the transmission
through the IGM. The models with sub-grid clumping more closely resemble
the highest redshift SDSS quasar spectra from Fan et al. (2006) than the calculations shown in Figure
\ref{fig:flux_4los}. We have also produced a model in which the entire volume is uniformly ionized with
$\avg{\Gamma_{\rm HI}} = 10^{-13} s^{-1}$, and $C_V = 10$ (not shown in plot), which has a nearly identical
average transmission profile to our $\avg{x_i} = 0.7$, $C_V = 10$ case. There is nothing
exotic about this model -- we expect this level of clumping from our semi-analytic calculations even
after reionization is complete. Our results suggest then that the $z \gtrsim 6.2$ 
Ly$\alpha$ proximity zone measurements are in fact compatible with a highly ionized IGM as previously
suggested by Bolton \& Haehnelt (2006) and Maselli et al. (2006). 

An important caveat to these results is that here we only
(crudely) model the enhanced recombination rate from sub-grid clumping; we ignore that our limited
simulation resolution likely underestimates the abundance of rare voids in the baryonic density distribution. 
This effect goes in the opposite direction of the recombination enhancement discussed here and leads to
an underestimate of the transmission. Interestingly, note that since the gas distribution is expected to become
more smoothed-out as time elapses after reionization, this effect -- while not necessarily the dominant
one -- tends to imply more absorption the {\em earlier} reionization occurs. We defer examining this to future
work.

Another interesting feature of our calculations is that the transmission
{\em close} to the quasar is sensitive to the level of clumpiness in the IGM. Sufficiently near 
the quasar, the
photoionization rate is dominated by the quasar itself, and any galactic component is essentially irrelevant
(Figure \ref{fig:1drt}). This means that transmission close to the quasar
provides a gauge of small-scale structure in the IGM that is insensitive to the details of the 
(highly uncertain) ionizing background (Cen \& Haiman 2000). Interestingly, the 
transmission profiles near
quasars appear to evolve
rapidly with redshift around $z \sim 6.2$ (Figure 14 of Fan et al. 2006), possibly a reflection of increased
clumpiness in the IGM at high redshifts. Since, given the quasar luminosity, one knows $\Gamma_{\rm HI}$ close
to the quasar, one might use this region to calibrate the relation between photoionization rate and transmitted flux.
In doing so, one may need to include the impact of quasar overdensities and infall, however (Faucher-Gigu\'ere et al. 2007), and account for uncertainties in the quasar ionizing luminosity
(Telfer et al. 2002).

\begin{figure}
\bc
\includegraphics[width=9.2cm]{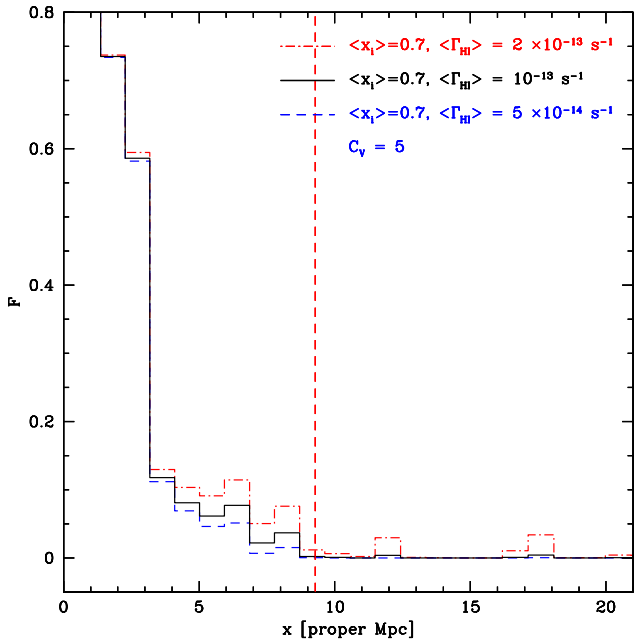}
\caption{Sensitivity of mock spectra to the volume-averaged, `background' hydrogen photoionization rate.
The precise value of the background hydrogen photoionization rate is affected by
several details of our modeling. The histograms show how mock absorption spectra vary in our 
$\avg{x_i} = 0.7$ case with varying $\avg{\Gamma_{\rm HI}}$, assuming no sub-grid clumping ($C_V = 5$).
The transmission profiles are impacted at relatively small distances ($r \sim 5$ proper Mpc) in
this example, owing to an upward fluctuation in the galactic photoionization rate.}
\label{fig:mock_v_gamma}
\ec
\end{figure}

As mentioned in \S \ref{sec:1d_rt}, even for a given model of bubble sizes, there is considerable
uncertainty in the precise galactic HI photoionization rate within the ionized bubbles (see
Equation \ref{eq:gammahi}).
This impacts the mock absorption spectra far from the quasar, where photoionization from galactic
sources is not overwhelmed by that from the quasar itself. This is in contrast to the influence of sub-grid
clumping which, at least in our simplified model, impacts the transmission even near the quasar. 
In order to illustrate this, we vary
the volume-averaged HI photoionization rate around our fiducial value, redo our 1D radiative
transfer calculations, and construct new mock spectra. The results of this calculation 
are shown in Figure \ref{fig:mock_v_gamma}. 

Indeed, boosting the mean intensity of the photoionizing
background by a factor of $\sim 2$ increases the transmission far from the quasar noticeably.
Note however that if the IGM is more clumpy than in our 
simulation, as we expect from our arguments above, a still
larger $\avg{\Gamma_{\rm HI}}$ may be required to achieve this much transmission. 
On the other hand, diminishing the background intensity by just a factor of $\sim 2$ basically
erases the transmission `spikes' far from the quasar. Note that this has little to do with the
proximity zone itself -- changing the photoionizing background does not impact absorption spectra
where the quasar itself dominates the photoionizing
flux -- and the effect is maximized far from the quasar where the photoionization from galactic
sources dominates. This just illustrates the usual sensitivity of absorption spectra to the
mean photoionization rate, except here we adopt a more realistic model for the {\em fluctuations}
in the photoionizing background. In principle, we can constrain our model for the ionizing sources,
bubble size distribution, and mean-free path to ionizing photons with these measurements: 
some of our models may overproduce
these transmission spikes compared to the data. In detail, this will require a better handle
on the gas density distribution in the IGM, and on our models for sub-grid clumping.

Based on our Figures \ref{fig:mock_v_clump} and \ref{fig:mock_v_gamma}, it seems plausible that the strong
redshift evolution in proximity zone sizes observed by Fan et al. (2006) (their Figure 14) near $z \sim 6.2$
reflects evolution in the level of small scale structure in the IGM, the abundance of rare voids in the gas density 
distribution, the intensity of the ionizing background,
and the cosmic mean gas density, rather than redshift evolution in the volume-weighted ionization
fraction. At this point our models are perhaps too crude to warrant a detailed comparison.
We intend to return to this issue in future work.

%%%%%%%%%%%%%%%%%%%%%%%%%%%%%%%%%%%%%%%%%%%%%%%%%%%%%%%%%%%%%%%%%%%%%%%%%%%%%%%
\subsection{The Impact of Lyman-limit Systems} \label{sec:lls}
%%%%%%%%%%%%%%%%%%%%%%%%%%%%%%%%%%%%%%%%%%%%%%%%%%%%%%%%%%%%%%%%%%%%%%%%%%%%%%%

In our previous calculations, quasar photons ionize gas at very large distances along some
lines of sight, $r \gtrsim 15-20$ proper Mpc. Should we really expect quasars to impact the ionization
state of gas at such large distances? The main concern here is that numerical simulations
under-produce Lyman limit systems owing to inadequate resolution and missing physics 
(Miralda-Escud\'e et al. 1996, Katz et al. 1996, Gardner et al. 1997, Meiksin \& White 2004, 
McDonald et al. 2005b, Kohler \& Gnedin 2007, Nagamine et al. 2004, 2007).
These studies have focused on $z \sim 3$, but this is likely even 
more of an 
issue at $z \sim 6$, especially if some `mini-halos' manage to survive 
photo-evaporative processes. Missing Lyman-limit systems may impact our results in several ways:
first, pre-quasar HII regions will be smaller at a given ionization fraction 
(Furlanetto \& Oh 2005, McQuinn et al. 2006b); second, quasar ionizing photons will propagate less
far; and finally, these dense absorbers may leave their imprint on Ly-$\alpha$ absorption spectra.
Note that these effects are not well-captured by the clumping-factor approach of the previous
section, which is more appropriate for `diffuse' gas in the IGM (Furlanetto \& Oh 2005, McQuinn et al.
2006a).

Let us begin by asking the question: how large should we expect the mean free path for ionizing
photons to be at $z \sim 6$? (For closely related calculations,
see the Appendix of Furlanetto \& Oh 2005.) Here, we
take an empirically-motivated approach and estimate the mean free path
of quasar ionizing photons from the observed column-density distribution in the $z \sim 3$ Ly-$\alpha$
forest, and extrapolate this result to higher redshifts. These quantities are very uncertain even at
$z \sim 3$. At $z \sim 3$ the essential difficulty is that absorption systems on the flat part of 
the curve of growth, where it is difficult
to accurately determine column densities, contribute significantly to
the effective optical depth for ionizing photons (see, e.g., Meiksin \&
Madau 1993). Owing to large amounts of saturated absorption in high redshift quasar spectra, 
the column density distribution is still more uncertain at high redshifts ($z \gtrsim 5$),
where Lyman-limit systems cannot be directly identified. Moreover, one might object
that the relevant calculation is the mean free path of ionizing 
photons {\it in the overdense environment
of a quasar host halo, with gas subjected to the enhanced ionization field from the quasar}, 
while 
we will calculate only the mean free path derived from column density distributions 
measured in {\it typical} regions of the IGM.

These objections are correct in detail, but our ensuing calculations should nonetheless provide 
useful estimates.
The mean free path can be calculated from the observed column
density distribution of Ly$\alpha$ absorbers, assuming they
are Poisson distributed. In this case, the effective optical
depth to photons at the ionization edge is given by (see e.g., Zuo \&
Phinney 1993):
\begin{eqnarray}
\tau_{\rm eff}(\lambda=912 \Ang, z_{\rm min}, z_{\rm max}) =
\int_{z_{\rm min}}^{z_{\rm max}} dz \times
\nonumber \\
\int dN_{\rm HI} f(N_{\rm HI}, z) \left(1 - e^{-N_{\rm HI} \sigma_{\rm HI}}\right).
\label{eq:taull}
\end{eqnarray}
The mean free path is then determined by the redshift extent, $z_{\rm
max} - z_{\rm min}$, over which $\tau_{\rm eff} = 1$. The quantity
$f(N_{\rm HI}, z)$ represents the column density distribution.
The mean free path for higher energy photons is naturally longer, although it scales
less rapidly with frequency than in a homogeneous IGM, where one expects $\lambda \propto \nu^3$.
For example, in the toy case
that the column density distribution follows a power law relation,
$f(N_{\rm HI}) \propto N_{\rm HI}^{-3/2}$, out to arbitrarily large column densities,
the mean free path scales as $\lambda \propto \nu^{3/2}$ (Zuo \& Phinney 1993).
Note that in Equation (\ref{eq:taull}) we include both the residual opacity from the diffuse, ionized
IGM (which is presumably incorporated in our calculations), and the opacity from Lyman-limit
systems, which is poorly-captured in our simulations.

\begin{figure}
\bc
\includegraphics[width=9.2cm]{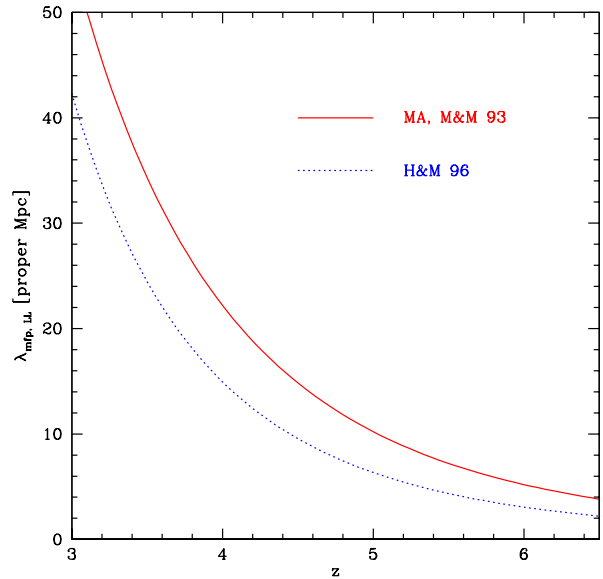}
\caption{Mean free path for hydrogen ionizing photons. The curves indicate the proper mean free
path of ionizing photons at the Lyman-limit frequency as a function of redshift, derived from two
different models for the column-density distribution in the Ly-$\alpha$ forest, extrapolated to high
redshift. The red curve adopts the Medium Attenuation model of Meiksin $\&$ Madau (1993), while the blue dotted
curve assumes the model of Haardt $\&$ Madau (1996). For both models, the mean free path of Lyman limit
photons is $r \lesssim 5$ proper Mpc by $z \sim 6$.
}
\label{fig:mfp}
\ec
\end{figure}
We calculate the mean free path from Equation (\ref{eq:taull}) using two different parameterizations
for the observed column density distribution. Specifically, we consider the Medium Attenuation model
of Meiksin \& Madau (1993), and the model from Haardt \& Madau (1996). The differences between
these models reflect uncertainties in measurements of the column density distribution. We simply
assume the redshift scalings in these parameterized models, based on observations at $z \sim 2-4$,
apply even out to $z \sim 6$.
The results of this 
calculation are shown in Figure \ref{fig:mfp}. 

In the Haardt \& Madau (1996) model, the mean free path for ionizing
photons at the Lyman-limit frequency is only $\lambda \sim 3$ proper Mpc at $z \sim 6$, and in
the Meiksin \& Madau (1993) model the mean free path is only $\lambda \sim 5$ proper Mpc by $z=6$.
Higher energy photons have longer mean free paths as described previously.
Note that Fan et al. (2002, 2006) estimate the mean free path at $z \sim 6$ directly from
absorption spectra data. These inferences are model dependent, but suggest still shorter mean free paths
than our extrapolations.
Low-column density systems and dense absorbers ($N_{\rm HI} \gtrsim 10^{17} \rm{cm}^{-2}$)
provide comparable contributions to the net opacity for Lyman-limit photons.
For comparison, 
quasar photons at the Lyman-limit in our 1D radiative transfer calculations have a 
mean free path of $\lambda \gtrsim 10$ proper Mpc (once the IGM is ionized -- i.e., this mean free
path is set by the opacity from residual neutral material along a sightline). 

These calculations
suggest that our 1D calculations overestimate the mean-free path of Lyman limit photons by a factor
of at least $\sim 2$. Note that a Lyman-limit system with $N_{\rm HI} \sim 10^{17} \rm{cm}^{-2}$, by 
definition, produces an opacity of only $\tau \sim 1$ and hence should attenuate impinging 
quasar flux, without entirely halting the quasar ionization front. Nevertheless, our estimates
suggest that even if the entire volume of the IGM is ionized, we may easily be 
overestimating the quasar flux at $r \sim 10$ proper Mpc by a factor of $e$ or $e^2$. As emphasized
previously, reducing the quasar flux by even a factor of a few should lead to significantly
more Ly-$\alpha$ absorption and make it harder to distinguish a partly neutral 
from a fully-ionized IGM on the
basis of Ly-$\alpha$ absorption spectra. Crossing a dense absorber will lead to a Lyman-limit
absorption system in a Ly-$\alpha$ forest spectrum, but this will be hard to distinguish 
sufficiently far from the quasar where the typical level of absorption is large.

%%%%%%%%%%%%%%%%%%%%%%%%%%%%%%%%%%%%%%%%%%%%%%%%%%%%%%%%%%%%%%%%%%%%%%%%%%%%%%%
\section{Conclusions} \label{sec:conclusions}
%%%%%%%%%%%%%%%%%%%%%%%%%%%%%%%%%%%%%%%%%%%%%%%%%%%%%%%%%%%%%%%%%%%%%%%%%%%%%%%

In this paper we argued that, even if the $z \sim 6$ SDSS quasars form when the IGM is partly
neutral, they are likely born into large HII regions (see also Yu \& Lu 2005). On spherical average, the 
ionization fraction
approaches the global mean after averaging over $R \sim 20-30$ co-moving Mpc/$h$. Nonetheless, much longer
skewers towards quasar host halos pass entirely, or predominantly, through ionized bubbles. 
These effects allow quasar fronts to travel {\em further} than expected in a uniformly-ionized medium
of the same ionized fraction. We find that the presence of large galaxy-created HII bubbles in patchy reionization models 
leads to a significant
sightline-to-sightline scatter in the extent of quasar ionization fronts. This complicates distinguishing
models on the basis of their observed proximity zones. Our results also contradict the
expectation for quasar fronts propagating into a homogeneous IGM, in which case front position scales
with neutral fraction simply as $R_s \propto X_{\rm HI}^{-1/3}$.

We emphasized that the resulting quasar absorption spectra are rather sensitive to uncertainties in the level of small
scale structure in the IGM, which impact the mapping between photoionization rate and transmission.
In accord with several previous studies 
(Fan et al. 2006, Bolton \& Haehnelt 2006, Maselli et al. 2006), we find that transmission profiles around the
$z \sim 6$ SDSS quasars do not generally reveal the extent of quasar ionization fronts. The transmission simply
falls close to zero when the combined radiation from the quasar and background galaxies is low enough -- 
essentially, the usual problem that small residual neutral fractions lead to complete absorption in Ly$\alpha$
at high redshift (see also Mesinger \& Haiman 2004). The precise neutral fraction at which 
complete absorption occurs, and the resulting 
absorption spectra, are sensitive to the 
level of small scale structure in
the IGM, the abundance of rare voids in the IGM, the mean-free path of ionizing photons in the IGM, and
the precise photoionization rate within ionized bubbles. 

Consequently, we encourage further theoretical study regarding the detailed gas density distribution in the IGM
(Miralda-Escud\'e et al. 2000) and sub-grid clumping factor prescriptions 
(e.g. Kohler et al. 2005, Trac \& Cen 2006). These models can be constrained by comparing with measurements
of the one-point probability distribution function in the $z \gtrsim 5$ Ly-$\alpha$ forest (Becker et al. 2006a), by 
examining the
region in the Ly$\alpha$ forest where the photoionization rate is dominated by the quasar 
itself (see \S \ref{sec:model_dep}, Cen \& Haiman 2000), and through other measures. These studies should be very useful for improved 
theoretical
modeling of $z \sim 6$ quasar proximity zones, and allow more definitive constraints on the volume-filling 
factor of HII
regions near $z \sim 6$. It might also be interesting to investigate Ly$\beta$ absorption in the proximity zone 
(Mesinger \& Haiman 2004, Bolton \& Haehnelt 2006), and other methods of identifying or
constraining the presence of damping wing absorption in the observed spectra 
(Mesinger \& Haiman 2006).

There are a few other additional uncertainties that warrant further attention. First, the quasar lifetime is very
uncertain observationally. Moreover, the `light bulb' model adopted here in which the 
quasar radiates at a fixed luminosity for
its entire lifetime is likely too simplistic (Hopkins et al. 2005, 2006). It might be interesting to use the $z \sim 6$ black-hole
growth simulations of Li et al. (2006), to provide a more realistic description of high redshift quasar life cycles.
Observationally, uncertainties in quasar redshifts may be significant (e.g. Shen et al. 2007) and impact proximity
zone measurements. Moreover, uncertainties from continuum-fitting -- particularly close to the quasar 
Ly$\alpha$ emission line -- warrant further investigation.

We mention several possible future research directions. First, many of the constraints on the $z \sim 6$ IGM adopt a 
simple model -- where one assumes a homogeneous photoionizing background, a uniform temperature-density 
relation, and that the baryonic distribution is uniformly Jeans-smoothed -- 
which, while remarkably successful at $z \sim 3$ 
(e.g. McDonald et al. 2005b, Viel \& Haehnelt 2006), should break down close to reionization. Demonstrating that this
model definitively fails at $z \sim 6$, and identifying which features of the model break, would
clearly strengthen the case for reionization activity near $z \sim 6$. On the other hand, the detailed
large scale reionization simulations that have been done to date and compared with quasar absorption spectra
(e.g. Kohler et al. 2005) have focused on only a single model. It is hard to determine how well alternate
models, in which reionization is more or less progressed, would match observations. 
Indeed, current measurements appear surprisingly
consistent with the simple assumption of a spatially uniform ionizing background (Lidz et al. 2006a, Liu et al. 2006).
It would be interesting to examine whether highly patchy reionization might overproduce the 
sightline-to-sightline scatter in the mean absorption. Also, there are certainly dark gaps in the spectra of
$z \sim 5.5$ quasars: are we sure of the conventional wisdom that the entire volume of the 
IGM is ionized by $z \sim 5.5$? The current line
of argument for $\avg{x_i} = 1$ at $z \sim 5.5$ comes from assuming a uniform $\Gamma_{\rm HI}$, and finding that 
a large
photoionization rate is required to match the mean transmission in the forest --  but this model may lead to
misleading conclusions, as one generally expects transmission in the large ionized bubbles that may exist before
reionization completes (Furlanetto et al. 2006b). The abundance of dark
gaps in $z \sim 6$ quasar spectra is a particularly interesting diagnostic for
future modeling (Fan et al. 2002, Lidz et al. 2002, Nusser et al. 2002, Gallerani et al. 2006, 
Fan et al. 2006).

We also intend to consider 21 cm imaging of quasar HII regions 
(Madau et al. 1997, Wyithe et al. 2005b). Here, one attempts to detect
quasar HII regions based on the brightness temperature contrast between
ionized gas around a known quasar and surrounding neutral gas. It would
be interesting to investigate whether surrounding galaxy-generated HII regions
complicate the morphology of these holes in the 21 cm sky.

Although we have focused on subtleties and challenges associated with interpreting quasar absorption spectra
at $z \sim 6$, we should reiterate that they currently provide our most extensive data set regarding the high
redshift IGM. Moreover, the rapid redshift evolution seen by Fan et al. (2006)
in most of the statistical properties of the $z \sim 6$ Ly$\alpha$ forest
is certainly striking, even if the precise theoretical implications are 
unclear. We believe that future modeling, along the lines mentioned above, should help us understand
the ionization state of the IGM at $z \sim 6$ and establish more definitive 
predictions for future surveys of the
high redshift IGM.

%%%%%%%%%%%%%%%%%%%%%%%%%%%%%%%%%%%%%%%%%%%%%%%%%%%%%%%%%%%%%%%%%%%%%%%%%%%%%%%
\section*{Acknowledgements}

We thank Zoltan Haiman, Andrei Mesinger, and Peng Oh for very helpful comments on a
draft.
We thank Oliver Zahn for useful discussions and for help with Figure 3.
We thank Mark Dijkstra, Claude-Andr\'e Faucher-Gigu\`ere, Steve Furlanetto, 
Simona Gallerani, and Yuexing Li for discussions on related issues.
The authors are supported by the David and Lucile Packard
Foundation, the Alfred P. Sloan Foundation, and NASA grants
AST-0506556 and NNG05GJ40G.

%%%%%%%%%%%%%%%%%%%%%%%%%%%%%%%%%%%%%%%%%%%%%%%%%%%%%%%%%%%%%%%%%%%%%%%%%%%%%%%

%%%%%%%%%%%%%%%%%%%%%%%%%%%%%%%%%%%%%%%%%%%%%%%%%%%%%%%%%%%%%%%%%%%%%%%%%%%

\end{document}